\documentclass[12pt]{iopart}

\usepackage{graphicx}

\begin{document}

\topical[Phenomenology of normal state transport in high-$T_c$ cuprates]{Phenomenology of the normal state in-plane
transport properties of high-$T_c$ cuprates}

\author{N E Hussey}

\address{H. H. Wills Physics Laboratory, Tyndall Avenue, Bristol BS8 1TL, UK}
\ead{n.e.hussey@bristol.ac.uk}

\begin{abstract}
In this article, I review progress towards an understanding of the normal state (in-plane) transport properties of
high-$T_c$ cuprates in the light of recent developments in both spectroscopic and transport measurement techniques.
Against a backdrop of mounting evidence for anisotropic single-particle lifetimes in cuprate superconductors, new
results have emerged that advocate similar momentum dependence in the transport decay rate $\Gamma$({\bf k}). In
addition, enhancement of the energy scale (up to the bare bandwidth) over which spectroscopic information on the
quasiparticle response can be obtained has led to the discovery of new, unforeseen features that surprisingly, may have
a significant bearing on the transport properties at the dc limit. With these two key developments in mind, I consider
here whether all the ingredients necessary for a complete phenomenological description of the anomalous normal state
transport properties of high-$T_c$ cuprates are now in place.

\end{abstract}


\submitto{\JPCM}
\maketitle

\section{Introduction}

As it enters its third decade, the field of high temperature superconductivity is finally leaving behind the heady days
of adolescence. As a sign of its growing maturity, it is slowly beginning to reveal more of its dark secrets under the
watchful gaze of dedicated experimentalists whose techniques have been honed almost uniquely in response to this, one
of the most profound problems in contemporary physics \cite{Sciencearticle}. Angle-resolved photoemission spectroscopy
(ARPES) and scanning tunnelling microscopy (STM) are arguably the two most high-profile examples of techniques whose
measurements on cuprates have undergone remarkable progress, but there are others, such as inelastic neutron scattering
(INS), that have enjoyed a similar transformation in recent years.

The improved momentum $k$ and energy $\epsilon$ resolution of both ARPES and STM (the latter achieving its
$k$-dependence via Fourier-transform scanning tunnelling spectroscopy) have opened up new vistas on the physics of the
in-plane quasiparticles that have revealed, amongst other discoveries, a remarkable \lq dichotomy' between the nodal
and anti-nodal eigenstates \cite{Zhou04, McElroy05}. This development has been supplemented by recent enhancement in
the energy ranges in both ARPES and INS over which the excitation spectra can be probed, up to energies of order the
bare bandwidth $W$ \cite{Graf07, Xie07, Valla07, Meevasana07, Chang07a} and the exchange interaction $J$
\cite{Hayden04, Tranquada04} respectively.

Within the transport community, the ability to resolve small spectral weight differences between the normal and
superconducting state (the so-called Ferrell-Glover-Tinkham sum rule) is testimony to the improvement in quality of
optical conductivity data (and its analysis) in recent years \cite{Molegraaf02, Santander-Syro03, Boris04}. This has
also led to a similar extension of the energy scale for probing the quasiparticle response \cite{vdM03} and the
development of methods to extract bosonic spectral densities across the full energy range \cite{Hwang07a}. Raman
spectroscopy is now providing $k$-resolution on scattering processes and excitations through careful polarization
selection \cite{DevereauxReview} whilst the application of angle-dependent magnetoresistance (ADMR) techniques to
overdoped cuprates has unravelled for the first time different components of the in-plane transport scattering rate
\cite{Majed06, Majed07, Analytis07}.

With this striking progress in the field over the last few years, it appears timely to take stock of the growing body
of experimental evidence and to assess to what extent our understanding of charge dynamics in the cuprates has
improved. In this topical review, I choose to focus on the normal state in-plane dc transport properties, resistivity,
Hall effect and magnetoresistance, but consider the output from a range of different experimental probes and how such
information might impact on our understanding of the cuprate transport problem.

The subject is important for a number of reasons. Just as in conventional superconductors, where the electron-phonon
scattering processes that dominate the electrical resistivity provided an important clue to the pairing interaction, so
an understanding of the normal state transport properties of high-$T_c$ cuprates (HTC) is widely regarded as a key step
towards the elucidation of the pairing mechanism for high temperature superconductivity. Whilst this remains the
ultimate goal, the anomalous transport behavior of the cuprates themselves has become arguably the most studied
phenomenon in the field of correlated electrons, a phenomenon that has inspired and engaged, at some point or other,
all but a fraction of the global condensed matter theoretical community \cite{Zaanen06}. And as experimentalists slowly
unravel more of the details, so the consensus of the theoretical community seems to be migrating away from
non-Fermi-liquid based models towards more Fermi-liquid-like \lq variations-on-a-theme'. Given this apparent paradigm
shift, it therefore seems prudent to examine to what extent conventional electronic band structure calculations and the
quasiparticle picture in general can account for the myriad of transport phenomena observed.

The article is arranged as follows. For the benefit of those new to the field, I begin by introducing the electronic
band structure of cuprates, before summarizing the key experimental observations, focusing on data that are generic to
most cuprates and for the most part, obtained on bulk single crystals. This is followed by a general discussion of the
current theoretical landscape and a review of recent transport and spectroscopic measurements that shed new light on an
old problem. I chose not to include detailed or comprehensive modelling in this paper; the task is simply to catalogue
those ingredients I believe to be necessary for a coherent phenomenological description of charge transport in HTC. As
this is a topical review, I focus on the most recent articles in the field and in acknowledging the vast amount of
literature that has helped to shape the problem over the past two decades, I apologize for their exclusion from this
article.

\section{Electronic structure}

\begin{figure}
\centering
\includegraphics[width=14.0cm,keepaspectratio=true]{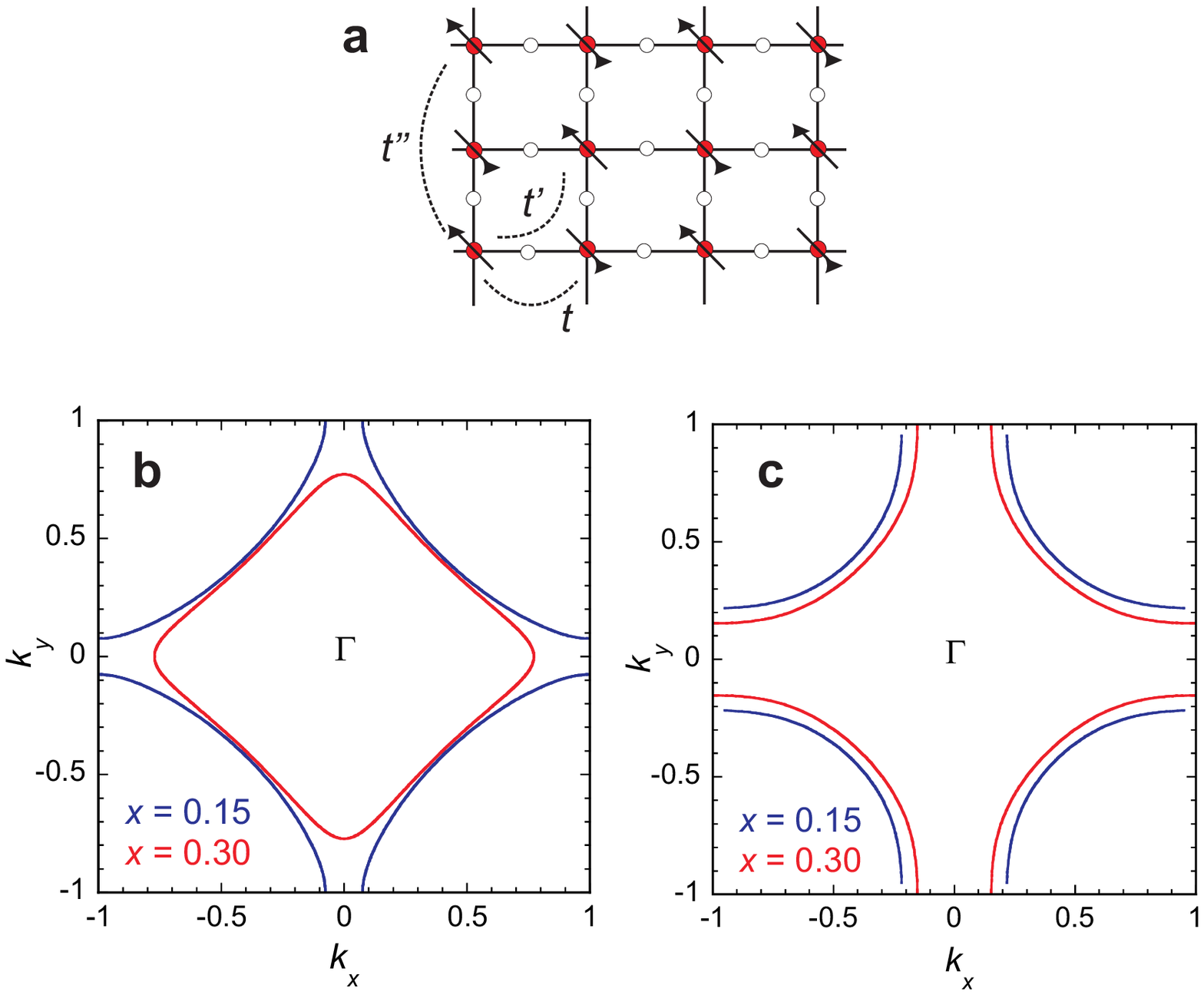} \caption{a) Schematic figure of the CuO$_2$
plane showing the spin alignments of the Cu spins at half-filling within the basal plane and the three principal
hopping parameters $t$, $t'$ and $t''$. b) Schematic 2D projection of the Fermi surface in La$_{2-x}$Sr$_{x}$CuO$_4$
for $p$ = 0.15 ($t'/t = 0.15$) and 0.30 ($t'/t = 0.12$). c) Similar projections for Tl$_2$Ba$_2$CuO$_{6+\delta}$ for
$p$ = 0.15 ($t'/t = 0.22$) and 0.30 ($t'/t = 0.22$). In all cases, $t''/t'$ = -0.5.} \label{Figure2}
\end{figure}

The structural element common to all HTC is the square planar CuO$_2$ plaquette shown schematically in Fig.
\ref{Figure2}a. Holes (or electrons) are doped onto the copper oxide planes from the charge reservoir layers located
between them. A combination of covalency, crystal field splitting and Jahn-Teller distortion then combine to create an
electronic structure whose highest partially-filled band has predominantly 3d$_{x^2-y^2}$ and O2p$_{x,y}$ character.
The resulting two-dimensional (2D) energy dispersion can be expressed in tight-binding representation as

\begin{equation}
\fl \epsilon({\bf k}) = \epsilon_0 - 2t(\cos k_x + \cos k_y) + 4t'(\cos k_x.\cos k_y) - 2t''(\cos
2k_x + \cos 2k_y) \label{eq1}\\
\end{equation}

At half-filling, with only nearest-neighbour ($t$) hopping, a diamond-like Fermi surface (FS) is expected. Inclusion of
next-nearest-neighbour ($t'$) hopping leads to a more rounded topology. Pavarini {\it et al.} identified an intriguing
correlation between $T_c$ and the ratio $t'/t$ for a large number of cuprate families \cite{Pavarini}. Low $T_c$
cuprates like La$_{2-x}$Sr$_{x}$CuO$_4$ (LSCO) and Bi$_2$Sr$_{2-x}$La$_x$CuO$_6$ (Bi-2201) have a relatively low
$t'/t$, whilst those with higher $T_c$ values, such as Bi$_2$Sr$_2$CaCu$_2$O$_{8+\delta}$ (Bi2212),
YBa$_2$Cu$_3$O$_{7-\delta}$ (YBCO) and Tl$_2$Ba$_2$CuO$_{6+\delta}$ (Tl2201), have much more rounded FS geometries
characteristic of the higher $t'/t$ values. These predictions have largely been verified by extensive ARPES
measurements \cite{Tanaka04}. The differences in topology are highlighted in Fig. \ref{Figure2}b and \ref{Figure2}c
where representative 2D FS projections of LSCO and Tl2201 respectively are shown for two different doping levels $p$ =
0.15 (near optimal doping) and 0.30 (beyond the superconducting dome). As one can see, the lower $t'/t$ values in LSCO
have the effect of driving the Fermi level $\epsilon_F$ below the van Hove singularity, inducing a crossover from a
hole-like to an electron-like FS, at a doping level of approximately $p$ = 0.18 \cite{Yoshida06}. In the other
cuprates, this does not occur, except perhaps for the anti-bonding band in Bi2212 \cite{Kaminski06}. As I shall
demonstrate here, these differences in FS topology are reflected in the dc transport properties.

Local density approximation (LDA) band structure calculations were largely ignored at the beginning of the HTC era due
to the failure of LDA to account for the antiferromagnetic insulating phase of the parent compound. However, recent
years have brought substantial improvement in theoretical techniques that include strong correlations in first
principles approaches. Moreover, a combination of new-generation ARPES \cite{ZhouReview} and ADMR \cite{Hussey03b} have
revealed that many details of the original LDA calculations are in fact reproduced experimentally, particularly in the
more highly doped cuprates ($p > 0.15$). The correct Fermiology of weakly doped cuprates, on the other hand, is still
to be resolved; whilst no-one disputes the existence of a (pseudo)gap in the normal state excitation spectrum, its
manifestation on the (remnant) FS and its evolution with temperature and doping remain highly controversial
\cite{Kanigel06, Doiron-Leyraud07}. In light of this, I will focus here more on the highly doped regions of the phase
diagram, though I shall return to discuss the situation in underdoped cuprates at the end.

\begin{figure}
\centering
\includegraphics[width=11.0cm,keepaspectratio=true]{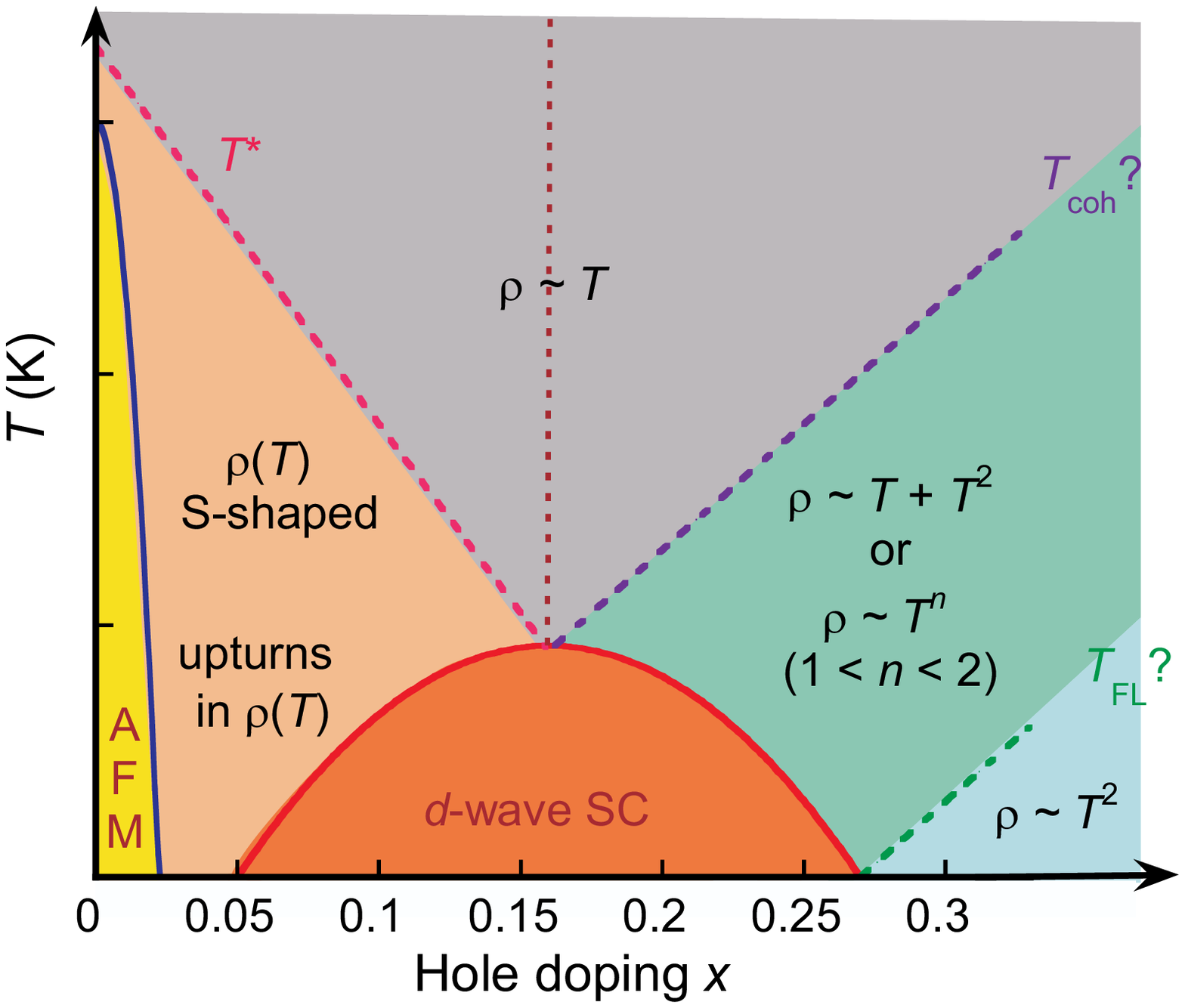} \caption{Phase diagram of (hole-doped)
cuprates mapped out in terms of the temperature and doping evolution of the in-plane resistivity $\rho_{ab}$($T$). The
solid lines are the phase boundaries between the normal state and the superconducting or antiferromagnetic ground
state. The dashed lines indicate (ill-defined) crossovers in $\rho_{ab}$($T$) behavior. The meanings of the labels
$T^*$, $T_{\rm coh}$ and $T_{\rm FL}$ are explained in the text.} \label{Figure1}
\end{figure}

\section{In-plane resistivity}

The in-plane resistivity $\rho_{ab}$($T$) of hole-doped HTC shows a very systematic evolution with doping that is
summarized in Fig.~\ref{Figure1}, where a schematic phase diagram of $p$-type cuprates is reproduced together with the
doping and temperature evolution of $\rho_{ab}$($T$). (Electron-doped cuprates will be dealt with at the end of this
section). The solid lines are the phase boundaries between the normal state and the superconducting or
antiferromagnetic ground state, whilst the dashed lines indicate (ill-defined) crossovers in $\rho_{ab}$($T$) behavior,
each of which, may or may not be associated with a fundamental change in the nature of the electronic states. Optimal
doping is indicated by the vertical dotted line corresponding to the pinnacle of the superconducting dome and the areas
to the left (right) of this line are the underdoped (overdoped) regions of the phase diagram respectively.

In the underdoped (UD) cuprates, $\rho_{ab}$($T$) varies approximately linearly with temperature at high $T$, but as
the temperature is lowered, $\rho_{ab}$($T$) deviates downward from linearity due to the partial removal of a dominant
scattering channel, as evidenced by the drop in 1/$\tau$($\omega$) seen in infrared spectroscopy \cite{TimuskStatt}.
This change of slope in $\rho_{ab}$($T$) was initially interpreted as a \lq kink' in $\rho_{ab}$($T$) at $T$ = $T^*$
(marked on the Figure) \cite{Bucher93, Ito93}. Plots of the derivative d$\rho_{ab}$/d$T$ showed however that
$\rho_{ab}$($T$) in fact first deviates from linearity at a much higher $T$ \cite{Hussey97}. Moreover, in the vicinity
of $T^*$, there is no additional feature in d$\rho_{ab}$/d$T$; the change of slope is a very gradual, continuous
process with no clear evidence of a phase transition below $T^*$. In the more anisotropic cuprates such as LSCO
\cite{Nakano98} and Bi-2212 \cite{Watanabe97}, it has proven difficult to distinguish between deviations from linearity
due to genuine pseudogap effects and those due to paraconductivity fluctuations near $T_c$. The dashed line depicting
$T^*$ in Fig. 1 reflects this ill-defined nature.

At sufficiently low temperatures, $\rho_{ab}$($T$) of UD cuprates develops an upturn, suggestive of some form of (as
yet unidentified) electronic localization. This upturn is characterized by a marked log($1/T$) dependence
\cite{Ando95}. The critical doping level $p_{\rm crit}$ at which these upturns occur differs amongst the various
cuprate families, being close to optimal doping in LSCO \cite{Boebinger96}, at 1/8 doping in La-doped Bi2201
\cite{Ono00} and close to the insulator/superconducting boundary in YBCO \cite{Proust}. This trend towards lower
$p_{\rm crit}$ with increasing purity suggests that the onset of localization is in fact disorder-driven.

Optimally-doped (OP) cuprates are characterized by a $T$-linear resistivity that survives for all $T > T_c$. Despite
the large variations in (optimal) $T_c$ and in the crystallography of individual cuprate families, $T$-linear
resistivity is a universal feature at optimal doping, confirming that it is intrinsic to the CuO$_2$ planes. Moreover,
as illustrated in Table \ref{table1}, the value of $\rho_{ab}$ at $T$ = 300K normalized to a single CuO$_2$ plane is
largely independent of the chemical composition of the charge transfer layers. By contrast, the electrical anisotropy
$\rho_c$/$\rho_{ab}$ varies by approximately 5 orders of magnitude between the different families of cuprates (see
Table 1). The $\rho_{ab}$ values themselves are large when compared with conventional superconductors. Given that the
dc resistivity (conductivity) depends on both the normal state plasma frequency $\Omega_{\rm pn}^2$ (= $ne^2/\epsilon_0
m^*$ in a Drude picture, where $n$ is the carrier density and $m^*$ the effective mass) and the transport scattering
rate 1/$\tau_{tr}$ = $\Gamma$, it has proved difficult to conclude whether these high values are due to a small
coherent spectral weight, i.e. a small number of carriers with long lifetimes, or a large number of heavily damped
qp's. Recent analysis of optical conductivity data however does seems to suggest the latter \cite{Homes04}.

\begin{table}[t]
\caption{\label{table1} $\rho_{ab}$(300K), normalized $\rho_{ab}$(300K) and $\rho_c$/$\rho_{ab}$($T_c$) values for some
optimally doped cuprates.} \vspace{0.3cm}
\begin{indented}
\lineup
\item[]\begin{tabular}{@{}*{7}{l}}
\br
Compound & $\rho_{\parallel}$ ($T$ = 300K) & $\rho_{\parallel}$/layer (300K) & $\rho_{\perp}$/$\rho_{\parallel}$ ($T_c$) \\
& \0\0\0($\mu\Omega$cm) & \0\0($\mu\Omega$cm) & \\
\mr
YBa$_2$Cu$_3$O$_{6.95}$ & \0\0\0290 \cite{Ando04} & \0\0\0\0580 & 3 10$^1$ \cite{Hussey99}\\
La$_{1.83}$Sr$_{0.17}$CuO$_4$ & \0\0\0420 \cite{Ando04} & \0\0\0\0420 & 3 10$^2$ \cite{Hussey98}\\
Bi$_2$Sr$_{1.61}$La$_{0.39}$CuO$_6$ & \0\0\0500 \cite{Ando04} & \0\0\0\0500 & 1 10$^6$ \cite{Ono03}\\
Bi$_2$Sr$_2$CaCu$_2$O$_{8+\delta}$ & \0\0\0280 \cite{Giura03} & \0\0\0\0560 & 1 $10^5$ \cite{Giura03}\\
Tl$_2$Ba$_2$CuO$_{6+\delta}$ & \0\0\0450 \cite{Tyler97} & \0\0\0\0450 & 2 10$^3$ \cite{Manako92}\\
\br
\end{tabular}
\end{indented}
\end{table}

In low-$T_c$ OP cuprates, where the superconductivity can be destroyed by large magnetic fields, the $T$-linear
$\rho_{ab}(T)$ has been found to cross over to a higher power $T$-dependence eventually levelling off at some finite
residual value \cite{Boebinger96, Ono00}. In higher-$T_c$ OP cuprates where this is not possible, a higher power
$T$-dependence as $T \rightarrow$ 0 can also be inferred from the fact that the slope of the $T$-linear resistivity
often extrapolates to a {\it negative} intercept \cite{Tyler97, Yoshida99}. Thus the $T$-linear resistivity observed in
OP cuprates does {\it not} extend down to the lowest temperatures, at least with the same slope. Moreover, Ando {\it et
al.} recently showed that the region of strict $T$-linearity (in the normal state) is rather narrow, concentrated at or
around optimal doping \cite{Ando04}. This confinement of the $T$-linear resistivity to a narrow composition range near
optimal doping is more suggestive of electron correlation effects than strong phonon interactions and is often regarded
as a signature of quantum criticality, as demonstrated in the heavy fermion compounds \cite{Custers03}.

On the overdoped (OD) side, $\rho_{ab}$($T$) contains a significant supralinear contribution that can be interpreted
either as a sum of two components, one $T$-linear, the other quadratic, or a single power law $T^n$ where $n$ varies
smoothly from 1 at optimal doping to 2 at the SC/non-SC boundary on the OD side \cite{Manako92, Mackenzie96a, Naqib03}.
At sufficiently high-$T$ however, $\rho_{ab}$($T$) becomes $T$-linear once more. This crossover temperature
\cite{Naqib03} is marked in Fig. \ref{Figure1} as a coherence temperature $T_{\rm coh}$, in line with the suggestion
from the ARPES community that the onset of $T$-linear resistivity coincides with the loss of the qp (coherence) peak in
the energy dispersion curves \cite{Kaminski03}.

The crossover to purely quadratic $\rho_{ab}$($T$), characteristic of a correlated Fermi-liquid (FL), is only observed
beyond the superconducting dome. In heavily OD, non-superconducting La$_{1.7}$Sr$_{0.3}$CuO$_4$ for example, the
low-$T$ resistivity in zero-field is found to be purely quadratic up to $T$ = 50K \cite{Nakamae03}. The dashed line
marked $T_{\rm FL}$ represents this crossover to strictly $T^2$ resistivity and whilst its nomenclature hints at
conventional FL behavior, quantum oscillations, the classic signature of a FL, have never been observed in this region
of the phase diagram. Significantly, Mackenzie {\it et al.} measured $\rho_{ab}(T)$ of heavily OD Tl2201 down to 0.1K,
suppressing $T_c$ (= 15K) with a large magnetic field, and still found evidence for a finite $T$-linear term coexisting
with this $T^2$ term and surviving into the $T$ = 0 limit \cite{Mackenzie96a}. Such behaviour is manifestly
non-Fermi-liquid-like.

\begin{figure}
\centering
\includegraphics[width=7.5cm,angle=270,keepaspectratio=true]{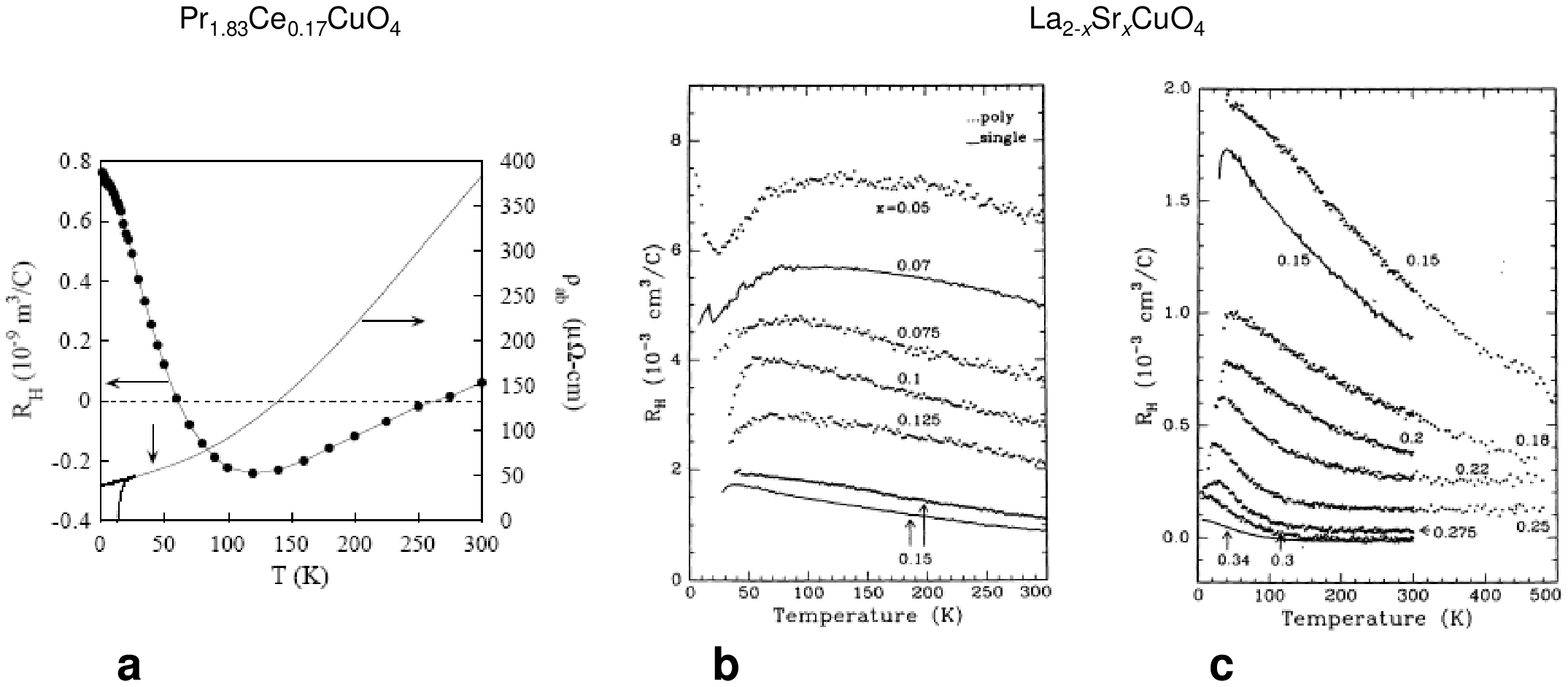} \caption{a) In-plane resistivity $\rho_{ab}(T)$
and Hall coefficient $R_{\rm H}(T)$ data for a slightly overdoped Pr$_{1.83}$Ce$_{0.17}$CuO$_4$ thin film. Reprinted
with kind permission from \cite{Fournier98}, figure 4. Copyright 1998 by the American Physical Society. b) $R_{\rm
H}(T)$ for underdoped La$_{2-x}$Sr$_x$CuO$_4$ \cite{Hwang94}. c) $R_{\rm H}(T)$ for overdoped La$_{2-x}$Sr$_x$CuO$_4$
\cite{Hwang94}. Reprinted with kind permission from \cite{Hwang94}, figure 4. Copyright 1994 by the American Physical
Society.} \label{Figure4}
\end{figure}

In marked contrast to what is observed in hole-doped cuprates, doping appears to have little or no effect on the
$T$-dependence of $\rho_{ab}$($T$) in $n$-type cuprates such as Pr$_{2-x}$Ce$_x$CuO$_{4+\delta}$ (PCCO), at least in
the intermediate temperature regime (100K $\leq T \leq$ 300K) \cite{Dagan07}. At low temperatures however,
$\rho_{ab}(T)$ becomes highly sensitive to changes in cerium and oxygen content with both metallic and insulating
behaviour being reported \cite{Fournier98, Dagan04, Gauthier07}. As shown in Fig. \ref{Figure4}a, a $T$-linear
resistivity is also observed in slightly overdoped PCCO between 30K and 40mK \cite{Fournier98}, reminiscent of what is
seen in OD Tl2201 \cite{Mackenzie96a}. Finally, a recent doping dependent study revealed that the limiting low-$T$ form
of $\rho_{ab}(T)$ could be expressed as $\rho_{ab}(T) = \rho_0 + AT^{\beta}$ with $\beta$ tending to unity at a
critical doping level $x_c$ = 0.165, again suggestive of a quantum critical point near optimal doping \cite{Dagan04}.

\section{In-plane Hall coefficient}

The in-plane Hall coefficient $R_{\rm H}$ of hole-doped cuprates varies markedly with both $p$ and $T$. This behaviour
is encapsulated by data on LSCO by Hwang {\it et al.} \cite{Hwang94}, which spans the entire (hole-doped) phase diagram
from the Mott insulator to the non-superconducting metal. These data are reproduced in Fig. \ref{Figure4}b and
\ref{Figure4}c for UD and OD LSCO respectively \cite{Hwang94}. At optimal doping ($x \sim 0.15$), $R_{\rm H}$ is found
to vary approximately as 1/$T$ over a wide temperature range \cite{Nishikawa94}, in apparent violation of conventional
Fermi-liquid theory.

According to band structure calculations, a \lq large' FS containing (1 + $x$) holes/Cu ion is predicted, centered
around the corners of the Brillouin zone as shown in Fig. \ref{Figure2}. In the UD region however, the carrier density
$n_{\rm H}$ at low $T$, estimated from the Drude relation $R_{\rm H} = 1/n_{\rm H}e$, approaches the \lq chemical' hole
concentration $x$ deduced from the formal valence of Cu$^{2+x}$ \cite{Ong87, Takagi89, Ando04b}. The observed scaling
of $R_{\rm H}$ with $x$ thus appears to suggest a violation of the Luttinger sum rule and either the presence of \lq
small' Fermi pockets containing $x$ holes or a series of Fermi arcs with an active (ungapped) arc length proportional
to $x$. At higher doping levels, $R_{\rm H}$ falls more rapidly (by 2 orders of magnitude for $0.05 < x < 0.25$
\cite{Ong87, Takagi89, Ando04b}) possibly reflecting the crossover from a small to a large FS. Finally for $x > 0.25$,
$R_{\rm H}$($T$) becomes negative at elevated $T$ \cite{Hwang94}.

Recent results from ARPES \cite{Yoshida06, Ino02, Yoshida03} support the doping evolution of the FS in LSCO inferred
from Hall measurements. At very low $x$, a weak but well-defined quasiparticle peak appears around the nodal point that
forms into an \lq arc' whose intensity increases smoothly with increasing $x$ \cite{Yoshida03} consistent with the
variation of the carrier number $n$. Beyond optimal doping, spectral intensity is strong everywhere on the FS
\cite{Yoshida06} and $\epsilon_F$ moves below the saddle near ($\pi$, 0), leading to a topological shift from a
hole-like, to an electron-like FS \cite{Ino02} as shown in Fig. \ref{Figure1}.

Despite this apparent consistency between the spectroscopic and transport probes, several outstanding issues remain, in
particular the magnitude and sign of $R_{\rm H}$ at low $T$. Whilst ARPES suggests that the FS in LSCO is electron-like
for $x \geq 0.18$, $R_{\rm H}$($T$$\rightarrow$0) remains positive at all doping levels up to $x$ = 0.34 \cite{Hwang94,
Tsukada06}. Given that they are, for the most part, single band metals, this sign problem is troublesome since at low
$T$, where scattering is dominated by impurities, the mean-free-path $\ell$ is expected to become independent of
momentum {\bf k}. In this so-called isotropic-$\ell$ regime, $R_{\rm H}$ (0) for a two-dimensional (2D) single-band
metal should reflect the sign of the dominant carrier, even if there are electron- and hole-like patches of FS
\cite{Ong91}, with a magnitude equal to the Drude result 1/$ne$.

At high $T$, $R_{\rm H}$ values in HTC agree roughly with LDA calculations. Thus, there appears to be a crossover from
the high-$T$ band-like regime to a low-$T$ anomalous regime. The crossover temperature $T_{\rm RH}$ systematically
increases with decreasing hole concentration \cite{Hwang94, Ong87}. A close correlation has been pointed out between
$T_{\rm RH}$ and the onset temperature of AFM correlations $T_{\chi}$, defined as a temperature where the uniform
susceptibility starts showing a rapid decrease, thereby implying a possible link between the AFM spin correlations and
the unusual behavior of $R_{\rm H}(T)$ \cite{Hwang94, Ong87}. Recent high temperature measurements of $R_{\rm H}$ in
underdoped LSCO by Ono and Ando \cite{OnoAndo07} however suggest that this strong $T$-dependence at lower doping levels
is in fact evidence for activated behaviour across the charge transfer gap that persists over a wide doping and
temperature range \cite{Uchida91}.

In electron doped cuprates, $R_{\rm H}(T)$ shows all the hallmarks of a system containing electron- {\it and} hole-like
carriers \cite{Wang91}. At low doping, $R_{\rm H}$ is negative and strongly $T$-dependent but around optimal doping,
$R_{\rm H}(T)$ changes sign, sometimes more than once, as the temperature is lowered. An example is shown in Fig.
\ref{Figure4}a for OP PCCO \cite{Fournier98}. In this doping region, the Hall resistivity $\rho_{xy}$ also displays
strong non-linearity with magnetic field $B$, and at certain temperatures, can even change sign with increasing $B$
\cite{Li07}. Finally, in the OD regime, $R_{\rm H}$ becomes positive and only weakly $T$-dependent. This evolution with
doping mirrors that of the Fermiology of $n$-type cuprates as deduced from ARPES. At low doping, the Fermi surface is
an electron pocket centred around ($\pi$, 0), with a volume proportional to $x$; the other regions in {\bf k}-space
being gapped presumably by spin-density-wave formation \cite{Matsui05, LinMillis05}. Upon further doping, spectral
weight eventually appears along the diagonal, giving rise to the two-carrier behaviour. Finally, as one moves beyond
optimal doping, the two sets of pockets merge into a large hole-like Fermi surface with a volume $\sim$ 1 - $x$
\cite{Armitage02}.

The relevance of the Hall coefficient as a gauge of carrier density and its evolution with temperature, particularly in
the hole-doped cuprates, was challenged by the discovery that the inverse Hall angle cot$\theta_{\rm H}$ (=
$\rho_{ab}/R_{\rm H}B$) had a unique and distinct $T$-dependence of its own \cite{Chien91}. In marked contrast to the
$T$-linear resistivity (at optimal doping), cot$\theta_{\rm H}(T)$ shows a quadratic $T$-dependence over a remarkably
broad temperature range. In OP LSCO, for example, cot$\theta_{\rm H}(T)$ $\sim A + BT^2$ between 50 K and 400K
\cite{Hwang94, OnoAndo07} whilst $\rho_{ab} \propto T$ up to 1000K \cite{GurvitchFiory}. This implicit \lq separation
of lifetimes' is a classic hallmark of the cuprates, and has led theorists to develop a number of radical ideas beyond
conventional Fermi-liquid theory, some of which will be outlined below.

Finally, whilst the $T^2$ dependence of cot$\theta_{\rm H}$ holds for a wide range of doping in most cuprates, it is
not the case for the Bi-based cuprates Bi2212 and Bi2201. In these systems, the power exponent of cot$\theta_{\rm
H}(T)$ is closer to 1.75 than 2 \cite{AndoMurayama, Konstantinovic00}. Detailed transport studies of both crystalline
and thin film samples of Bi2212 and Bi2201 have shown in fact that cot$\theta_{\rm H}(T) \sim A + BT^\alpha$  with
$\alpha$ steadily decreasing from $\sim 2$ to $\sim 1.6 - 1.7$ as one moves from the UD to the OD regime
\cite{AndoMurayama, Konstantinovic00}. This variable power law behavior in cot$\theta_{\rm H}(T)$ reveals a high level
of complexity in the phenomenology of normal state transport in HTC that has yet to be properly addressed.

\section{In-plane magnetoresistance}

\begin{figure}
\centering
\includegraphics[width=7.5cm,angle=270,keepaspectratio=true]{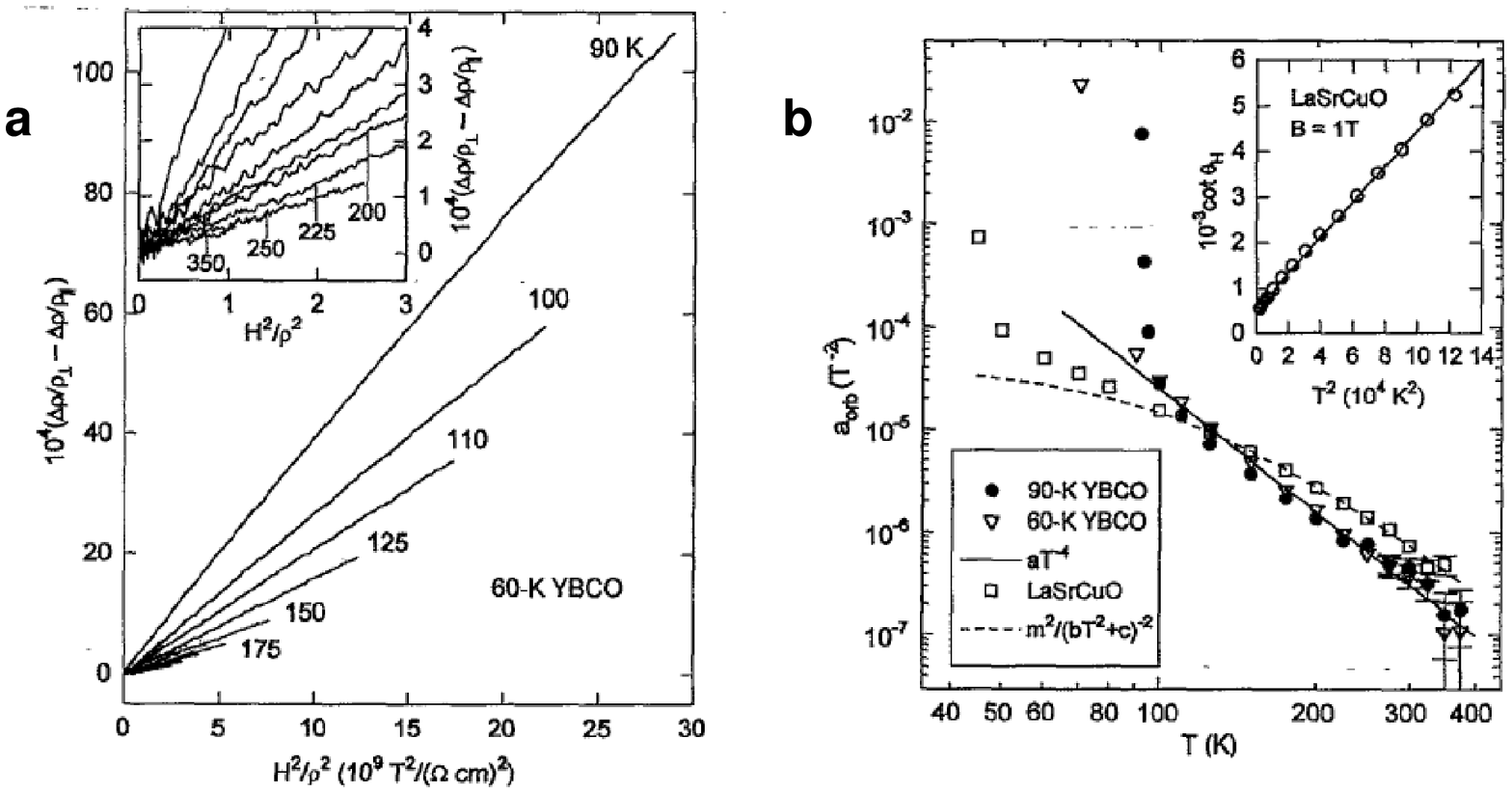} \caption{a) Kohler plot
for underdoped YBa$_2$Cu$_3$O$_{6.6}$ at intermediate (main) and high (inset) temperatures \cite{Harris95}. Reprinted
with kind permission from \cite{Harris95}, figure 4. Copyright 1995 by the American Physical Society. b) Temperature
dependence of the orbital part of the magnetoresistance in YBa$_2$Cu$_3$O$_{6.6}$, optimally doped YBa$_2$Cu$_3$O$_7$
and La$_{1.85}$Sr$_{0.15}$CuO$_4$ \cite{Harris95}. The inset shows the inverse Hall angle cot$\theta_{\rm H}$ vs $T^2$
in OP LSCO. Reprinted with kind permission from \cite{Harris95}, figure 3. Copyright 1995 by the American Physical
Society.} \label{Figure5}
\end{figure}

According to Boltzmann transport theory, the orbital transverse magnetoresistance (MR) of a metal $\Delta \rho/\rho
\propto (\omega_c \tau_{tr})^2$ where $\omega_c$ is the cyclotron frequency and $\tau_{tr}$ the transport lifetime. If
the only effect of a change of temperature or of a change of purity of the metal is to alter $\tau_{tr}$({\bf k}) to
$\lambda\tau_{tr}$({\bf k}) where $\lambda$ is not a function of {\bf k}, then $\Delta \rho/\rho$ is unchanged if $B$
is changed to $B/\lambda$. Thus the product $\Delta \rho \cdot \rho$ $(= \Delta \rho/\rho \cdot \rho^2)$ is independent
of $\tau_{tr}$ and a plot of $\Delta \rho/\rho$ versus ($B/\rho)^2$ is expected to fall on a straight line with a slope
that is independent of $T$ (provided the carrier concentration remains constant \cite{LuoMiley}). This relation, known
as Kohler's rule, is obeyed in a large number of standard metals, including those with two types of carriers, provided
that changes in temperature or purity simply alter $\tau_{tr}$({\bf k}) by the same factor. In HTC however,
conventional Kohler's rule is strongly violated; instead of the data collapsing onto a single curve, there is a marked
increase in the slope with decreasing temperature, as illustrated in the left panel of Fig. \ref{Figure5} for UD
YBa$_2$Cu$_3$O$_{6.6}$ \cite{Harris95}. Remarkably, this progression continues up to 350K (see inset).

Progress towards understanding this anomalous behavior came in the form of a modified Kohler's rule suggested by Ong
and co-workers \cite{Harris95}. They found that the in-plane orbital MR $\Delta \rho_{ab}/\rho_{ab}(T)$ follows the
$T$-dependence of tan$^2\theta_{\rm H}$ in both YBCO and LSCO much more closely than $\rho_{ab}^2$. The right panel of
Fig. \ref{Figure5} shows the temperature dependence of the orbital MR in UD YBCO, OP YBCO and OP LSCO. The solid and
dashed lines represent the $T$-dependence of tan$^2\theta_{\rm H}$ for both OP samples. Similar scaling was also
reported in OD Tl2201 \cite{Hussey96}. Intriguingly, only in OD non-superconducting LSCO is conventional Kohler's
scaling seemingly recovered \cite{Kimura96}.

Finally, in analyzing the normal state orbital MR, one must not overlook contributions to the orbital MR from
paraconductivity terms which can influence the in-plane magnetotransport over a wide temperature range in HTC due to
their small superconducting coherence length and strong two-dimensionality. In highly anisotropic Bi2212 for example,
apparent Kohler's law violation up to 300 K can be attributed almost entirely to superconducting fluctuations
\cite{Latyshev} whilst in UD YBCO, fluctuation contributions are believed to persist up to temperatures of order $T^*$
\cite{Leridon01}. For a comprehensive review of fluctuation effects in HTC, please refer to Ref.\cite{LarkinVarlamov}.

\section{The theoretical landscape}

The origin of the strong $T$-dependence of $R_{\rm H}$ and its large magnitude at low doping have been the subjects of
intense debate within the community. In a simple Drude picture, the sharp rise in $R_{\rm H}(T)$ suggests a loss of
carriers with decreasing temperature, due perhaps, to the opening of the pseudogap \cite{LuoMiley, AlexandrovMott} or
proximity to a vHs \cite{BokBouvier}. From this perspective, the non-monotonic $R_{\rm H}(T)$ observed in most cuprates
at low doping (see, e.g. Fig. \ref{Figure4}b) ought only be interpreted as a repopulation of states. This conclusion
however is not supported by magnetic susceptibility or specific heat data that show a suppression of low energy states
with decreasing temperature \cite{Cooper96}. In a conventional metal, a strong $T$-dependence of $R_{\rm H}$ can also
arise due to multiple band effects. While this is evident in the electron-doped cuprates \cite{Fournier98} and other
multiple band quasi-2D metals such as Sr$_2$RuO$_4$ \cite{Mackenzie96b}, it is unlikely to be applicable to the
majority of hole-doped cuprates where the transport is dominated by a single band.

Attempts to explain the anomalous behavior of $\rho_{ab}(T)$ and $R_{\rm H}(T)$ in cuprates within a FL scenario have
thus centered around the assumption of a (single) transport scattering rate whose magnitude varies around the in-plane
FS. In-plane anisotropy in 1/$\tau_{tr}$ has been attributed to anisotropic e-e (Umklapp) scattering \cite{Hussey03} as
well as to coupling to a singular bosonic mode; be it spin fluctuations \cite{Carrington92, MonthouxPines}, charge
fluctuations \cite{Castellani95}, $d$-wave superconducting fluctuations \cite{IoffeMillis} or more recently,
Pomeranchuk fluctuations \cite{Dell'Anna07}. Generating a clear \lq separation of lifetimes' within such single
lifetime scenarios however has proved very difficult, requiring as it does a very subtle balancing act between
different regions in $k$-space with distinct $T$-dependencies. In the \lq cold spots' model of Ioffe and Millis for
example, the phenomenological scattering rate contains two terms, an isotropic FL scattering rate 1/$\tau_{\rm FL} \sim
T^2$ and a $T$-independent scattering rate 1/$\tau_0$ that is large everywhere except the nodal directions
\cite{IoffeMillis}. Whilst this model correctly explains the $T$-linear resistivity and quadratic Hall angle, the
anisotropy required to separate the transport and Hall lifetimes gives rise to an orbital MR that is one order of
magnitude too large and has a much stronger $T$-dependence than is observed. These discrepancies can be resolved by the
introduction of a \lq shunt' scattering rate maximum $\Gamma_{\rm max}$ which acts to reduce the overall effective
anisotropy \cite{Hussey03}. A modified Kohler's rule of the correct magnitude is then reproduced but again its success
relies heavily on a subtle balancing of anisotropies in the elastic and inelastic channels.

Given this reliance on detail, other more exotic models, based on non-FL physics, have gained prominence within the
community; most notably the two-lifetime picture of Anderson \cite{Anderson91} and the Marginal Fermi-liquid (MFL)
phenomenology of Varma and co-workers \cite{Varma89}. In the two-lifetime approach, scattering processes involving
momentum transfer perpendicular and parallel to the FS are governed by independent transport and Hall scattering rates
1/$\tau_{tr}$ ($\propto T$) and 1/$\tau_{\rm H}$ ($\propto T^2$). In conventional FL's of course, $\tau_{tr}$ is equal
to $\tau_{\rm H}$. Allowing $\tau_{\rm H}$ to be independent of $\tau_{tr}$, the inverse Hall angle can now be written
as cot$\theta_{\rm H} = \sigma_{xx}/\sigma_{xy} \propto 1/\tau_{\rm H}$. Thus the different behavior of $\rho_{ab}(T)$
and cot$\theta_{\rm H}(T)$ reflects the different $T$ dependencies of 1/$\tau_{tr}$ and 1/$\tau_{\rm H}$. The
enhancement of $R_{\rm H}$ then comes from the fact that $\tau_{\rm H}$ becomes larger than $\tau_{tr}$ at low $T$.
This model received strong support from Hall measurements on Zn-doped YBCO that showed cot$\theta_{\rm H} = A + BT^2$
to be robust to Zn doping with $B$ remaining constant and $A$ increasing in proportion to the Zn concentration
\cite{Chien91}, suggesting that the $T^2$ (inverse) Hall angle is a well defined fundamental quantity representing
1/$\tau_{\rm H}$. Later measurements on Co-doped YBCO \cite{Carrington92} together with MR measurements on YBCO and
LSCO \cite{Harris95} appeared to affirm the robustness of 1/$\tau_{\rm H}$. In particular, the $T$-dependence of the MR
could be quantitatively explained by modifying Kohler's rule such that $\rho_{ab}/\rho_{ab} \propto (\omega_c.\tau_{\rm
H})^2 \propto 1/(A + BT^2)^2$ and consequently ($\rho_{ab}/\rho_{ab})$/tan$^2\theta _{\rm H}$ became a constant.

Whilst the two lifetime model of Anderson and co-workers has been successful in reproducing the experimental situation
in OP cuprates, it does not appear to be consistent with ARPES results and is yet to explain the evolution of the
transport phenomena across the full HTC phase diagram. The MFL hypothesis argues that optimum $T_c$ lies in proximity
to a quantum critical point and as a result, qp weight vanishes logarithmically at the FS with the corresponding
imaginary part of the self-energy governed simply by the temperature scale \cite{Varma89}. In contrast to the
two-lifetime picture, MFL theory assumes a single $T$-linear scattering rate but introduces an unconventional expansion
in the magnetotransport response. The Hall angle, for example, is given by the square of the transport lifetime
\cite{VarmaAbrahams}, an idea that has received empirical support from infrared optical Hall angle studies
\cite{Grayson02}. In order to account for the observed magnetotransport behavior in cuprates, Varma and Abrahams
introduced anisotropy into their MFL phenomenology via the {\it elastic} (impurity) scattering rate by assuming small
angle scattering off impurities located away from the CuO$_2$ plane \cite{VarmaAbrahams}. Whilst this hypothesis seems
consistent with certain ARPES measurements \cite{Valla00} {\it and} transport measurements \cite{Narduzzo07} (see
below), the legitimacy of the expansion in small scattering angle used in Ref. \cite{VarmaAbrahams} has been
subsequently challenged \cite{Hlubina01, Carter02}. In particular, it has been argued that the conditions that lead to
a separation in lifetimes do {\it not} reproduce the violation of Kohler's rule \cite{Carter02}. Moreover, although the
predictions of MFL theory appear compatible with the empirical situation in OP cuprates, their applicability to the
rest of the cuprate phase diagram is less evident. In particular, the gradual convergence of the $T$-dependencies of
$\rho_{ab}(T)$ and cot$\theta_{\rm H}(T)$ in OD cuprates sits uncomfortably with the idea of $\tau_{\rm H}$ scaling
with the square of the transport lifetime.

Thus, at the time of writing, none of the leading proposals appear to stand up to close scrutiny with the full
complement of experimental data. It appears there is still some way to go before a coherent theoretical description of
transport in HTC can emerge and if the truth be told, little progress has been made over the last few years. On the
experimental side however, there may been a number of recent investigations that could ultimately lead to major
advances in our comprehension of the origin of the anomalous transport properties and of the character of the qp
scattering, both as a function of momentum and energy. In the following sections, I summarize these results, starting
with evidence for anisotropic scattering, and discuss the impact of such findings on our interpretation of the dc
normal state transport.

\section{Anisotropic quasiparticle scattering}

Evidence for $T$-dependent basal-plane anisotropy in the transport scattering rate in HTC first came from studies of
the angle-dependent interlayer magnetoresistance $\Delta \rho_c/\rho_c$ in OD Tl2201 \cite{Hussey96}. On rotating the
magnetic field $B$ within the basal plane, $\Delta \rho_c/\rho_c$ was found to exhibit four-fold anisotropy with an
amplitude that scaled as $B^4$ in accordance with Boltzmann transport analysis for an anisotropic quasi-2D FS. The
ratio of the amplitude of these four-fold oscillations to the size of the higher order $B^4$ term in the isotropic MR
showed a significant $T$-dependence that corresponded to an {\it increasing} anisotropy of the in-plane scattering rate
(or more precisely, 1/$\ell$) with increasing $T$, whilst the sign of the four-fold term indicated that the most
intense scattering occurred near the saddles.

Subsequent ARPES measurements have revealed more information about the nature of the electronic states at different
loci on the 2D FS and the corresponding self-energy corrections $\Sigma$ = Re$\Sigma$ + Im$\Sigma$ \cite{Damascelli}.
Whilst there remains some controversy regarding the precise energy, momentum and doping dependence of the self-energy,
it is now well established that basal plane anisotropy in the \lq single-particle' scattering rate Im$\Sigma$ does
exist, with ill-defined qp lineshapes (in the normal state) at the so-called \lq anti-nodal' points near ($\pi$, 0)
co-existing with sharper qp peaks along ($\pi$, $\pi$).

Unfortunately, analysis of the energy dependence of Im$\Sigma$ is often complicated by the onset of superconductivity,
particularly since the reduction in scattering below $T_c$ occurs not only in the inelastic, but also in the elastic
channel, giving rise to additional energy-dependent terms in the self-energy \cite{Zhu04}. For this reason, I consider
here only data reported in the normal state ($T > T_c$) or for energies $\omega > \Delta$, the maximum superconducting
gap value. One further advantage of focusing on the normal state is that any kink features between 50 and 100 meV, that
tend to complicate the energy dependence of Im$\Sigma$, are relatively weak there.

Early measurements on OP Bi2212 showed evidence for MFL behaviour of the normal state ARPES lineshape along the nodal
direction with Im$\Sigma \propto \omega$ \cite{Valla99}.  More recent data however appear to suggest a dominant
contribution to Im$\Sigma$ that is {\it quadratic} in $|\omega |$ at low energies \cite{Kordyuk04, Koralek06}, crossing
over to a $\omega$-linear dependence only above 50 - 100 meV. As one moves into the OD region, Im$\Sigma$ along ($\pi,
\pi$) becomes purely quadratic (at least up to 100 meV) \cite{Kordyuk04}. Away from the nodes, it has proved difficult
to determine the precise form of Im$\Sigma(\omega)$ since the lineshapes are invariably broad and the qp states
incoherent, due presumably to a strong dressing or scattering of the charge carriers in this region in {\bf k}-space
\cite{Borisenko03, Cuk04, JAW}. Kaminski {\it et al.} showed in fact that the qp states near ($\pi$, 0) become
incoherent above a certain doping-dependent temperature which coincides with the onset of $T$-linear resistivity and is
labelled $T_{\rm coh}$ in Fig. \ref{Figure1} \cite{Kaminski03}. Along the nodes meanwhile, the qp states appear to
remain coherent to much higher temperatures \cite{Koralek06}. Finally, by carefully scanning the intermediate regions
along the FS, Chang {\it et al.} have recently uncovered a strong, highly anisotropic $\omega$-linear contribution in
OP LSCO that becomes negligible as one approaches ($\pi, \pi$) \cite{Chang07b}.

Extrapolating back to zero energy or to zero temperature, we find evidence for significant anisotropy in the
single-particle {\it elastic} scattering rate or inverse mean-free-path (as determined by the width of the momentum
distribution curve) in Bi2212 \cite{Valla00}, Bi2201 \cite{Kondo06} and LSCO \cite{Yoshida07, Chang07b}. In all cases,
the maximum is located near the saddles where the density of states is largest, consistent with the picture of Abrahams
and Varma for small-angle scattering off out-of-plane impurities \cite{VarmaAbrahams, Abrahams00}.

Given this mounting evidence for anisotropic single-particle lifetimes in HTC, let us now turn to consider the
transport scattering rate 1/$\tau_{\rm tr}$. In a 2D metal with an anisotropic $\ell$({\bf k}), Ong showed that
$\sigma_{xy}$ is determined by the area (curl) swept out by $\ell$({\bf k}) as it is traced around the FS \cite{Ong91}.
(Ong referred to this as the \lq Stokes area'). Thus, anisotropy in $\ell$({\bf k}), in addition to the local FS
curvature, plays a fundamental role in determining both the magnitude and sign of the Hall voltage in 2D metals.
Accordingly, cot$\theta_{\rm H}(T)$ will be dominated by those regions of the FS where the curvature is greatest and/or
where the qp mean-free-path is longest.

\begin{figure}
\centering
\includegraphics[width=5.5cm,angle=270,keepaspectratio=true]{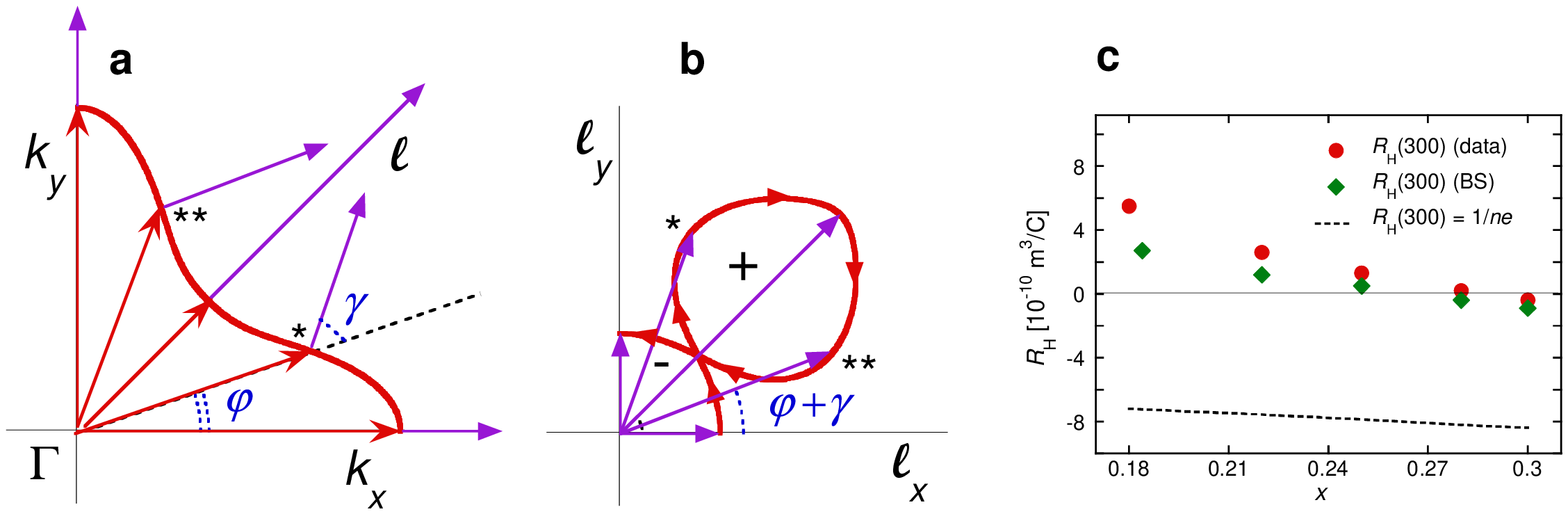}
\caption{(a) Section of 2D Fermi surface with pronounced negative curvature. The purple (red) arrows indicate the
direction and length of $\ell$($\varphi$) ($k_F(\varphi$)), as explained in the text. b) Polar plot of $\ell$$(\varphi
+\gamma)$. The red arrows indicate the circulation of each loop and the -/+ signs indicate the corresponding sign of
$\sigma_{xy}$. The resultant $\sigma_{xy}$ is determined by the difference in the areas of the two counter-rotating
loops ($\times 4$) \cite{Ong91}. c) Measured (red circles, \cite{Hwang94}) and band-derived (green diamonds) $R_{\rm
H}$(300) values in overdoped LSCO compared to the Drude value 1/$ne$ (dashed line).} \label{Figure6}
\end{figure}

The Ong construction is illustrated schematically in Fig.~\ref{Figure6}. The solid red line in Fig.~\ref{Figure6}a
represents a 2D-projected FS, similar to that shown in Fig.~\ref{Figure2}b for OD LSCO but with exaggerated negative
curvature. This FS geometry gives rise to alternating sectors on the FS that have electron- and hole-like character.
The purple arrows indicate the direction and length of the ${\bf \ell}$-vector for selected points on the FS. The
angles between ${\bf \ell}$ and {\bf k} and between {\bf k} and the $k_x$ axis are labelled $\gamma$ and $\varphi$
respectively. As {\bf k} moves along the FS away from the $k_x$ axis, $\varphi$ and $\gamma$ increase in the same
sense. At a particular {\bf k}-point, marked by $*$, $\kappa$ = d$\gamma$/d$\varphi$ changes sign and remains negative
until $\varphi$ reaches $**$. At $\varphi = \pi/2$, $\gamma$ is once again equal to zero. Were $\ell$({\bf k}) to be
isotropic, the corresponding \lq $\ell$-curve' would be a circle and $R_{\rm H}$ would be determined simply by the FS
area with a sign that reflects its location in the Brillouin zone. If ${\bf \ell}$ is anisotropic however, as shown in
Fig.~\ref{Figure6}b, loops of different circulation will appear in the $\ell_x$-$\ell_y$ plane. Ong demonstrated that
$\sigma_{xy}$ will then be determined by the sum of the areas enclosed by the primary (negative) and secondary
(positive contribution to $\sigma_{xy}$) loops. In the simulation shown in Fig.~\ref{Figure6}b, the secondary loops
have the largest area and so $R_{\rm H}$ is of opposite sign.

In OD LSCO ($x \geq$ 0.18), the FS encircles the $\Gamma$ point in the Brillouin zone (see Fig.~\ref{Figure2}b) and has
similar, if less pronounced, negative curvature to that shown in Fig.~\ref{Figure6}a \cite{Yoshida06}. The dashed line
in Fig.~\ref{Figure6}c represents $R_{\rm H}$ calculated using the Luttinger sum rule and the isotropic-$\ell$
approximation, whilst the red circles are experimental $R_{\rm H}$ values at $T$ = 300 K for different Sr contents
\cite{Hwang94}. For all but the highest doping level, $R_{\rm H}$(300) is found to be of opposite sign to the expected
Drude result with a magnitude that is significantly smaller than 1/$ne$. At high temperatures, where scattering is
sufficiently intense, one expects $\tau_{tr}$ to be isotropic within the plane. In the absence of any experimental
evidence for FS reconstruction in OD LSCO \cite{Yoshida06}, interpretation of these marked discrepancies in $R_{\rm
H}$(300) within a band picture must therefore imply strong in-plane anisotropy in the Fermi velocity $v_F(\varphi)$. As
it happens, this is precisely what is observed experimentally. For $x$ = 0.30, for example, $v_F$ near the saddle
points is 3-4 times smaller than along the nodal directions \cite{Yoshida07}. This anisotropy, coupled with the FS
curvature, accounts almost entirely for the deviation of $R_{\rm H}$ at $T$ = 300 K from its isotropic-$\ell$ limit, as
indicated by the green triangles in Fig. \ref{Figure6}c \cite{Narduzzo07}. As $x$ decreases, the FS retreats towards
the vHs, causing the band anisotropy to grow and correspondingly, $R_{\rm H}$(300) to deviate even further from the
isotropic-$\ell$ limit until eventually the vHs is crossed around $x$ = 0.18 \cite{Yoshida07}.

Whilst band anisotropy can appear to account for the anomalous $R_{\rm H}$ values at $T$ = 300K, it cannot explain the
$T$-dependence. For all $x \geq$ 0.18, $R_{\rm H}$($T$) is seen to become increasingly {\it more} positive with
decreasing temperature (see Fig.~\ref{Figure4}c), implying that anisotropy in $\ell({\bf k})$ becomes even more
pronounced as one approaches the elastic limit. In order to account for this anomalous $T$ dependence, additional
anisotropy in the {\it elastic} scattering rate $\Gamma_0(\varphi)$, presumably due to static impurities, is required.
In the isostructural ruthenate compound Sr$_2$RuO$_4$, the isotropic-$\ell$ approximation is found to be obeyed at low
$T$ \cite{Mackenzie96b}. La doping for Sr introduces disorder between the RuO$_2$ planes, and whilst it has a
negligible effect on the de Haas-van Alphen frequencies (and hence the FS volume), it induces both a magnitude {\it
and} a sign change in the zero-temperature Hall coefficient $R_{\rm H}$(0) \cite{Kikugawa}. As alluded to earlier,
anisotropy in $\Gamma_0(\varphi)$ can arise due to small-angle scattering off dopant impurities located between the
conducting planes \cite{VarmaAbrahams}. If $d$ is the characteristic distance of such dopants from a plane, the
electron scattering will involve only small momentum transfers $\delta k \leq d^{-1}$. Then, $\Gamma_0$($\varphi$) is
proportional to $\delta k$ and the local density of states, i.e. to 1/$v_F(\varphi)$. A predominance of forward
impurity scattering in cuprates has also been invoked to explain the weak suppression of $T_c$ with disorder
{\cite{Lee01}, and the energy and $T$-dependence of the single-particle scattering rate Im$\Sigma$ below $T_c$
\cite{Zhu04}. By introducing anisotropy of this form in $\Gamma_0(\varphi)$, the full $T$-dependence of $R_{\rm
H}$($T$) in OD LSCO can then be qualitatively and quantitatively explained \cite{Narduzzo07}.

\begin{figure}
\centering
\includegraphics[width=12.0cm,keepaspectratio=true]{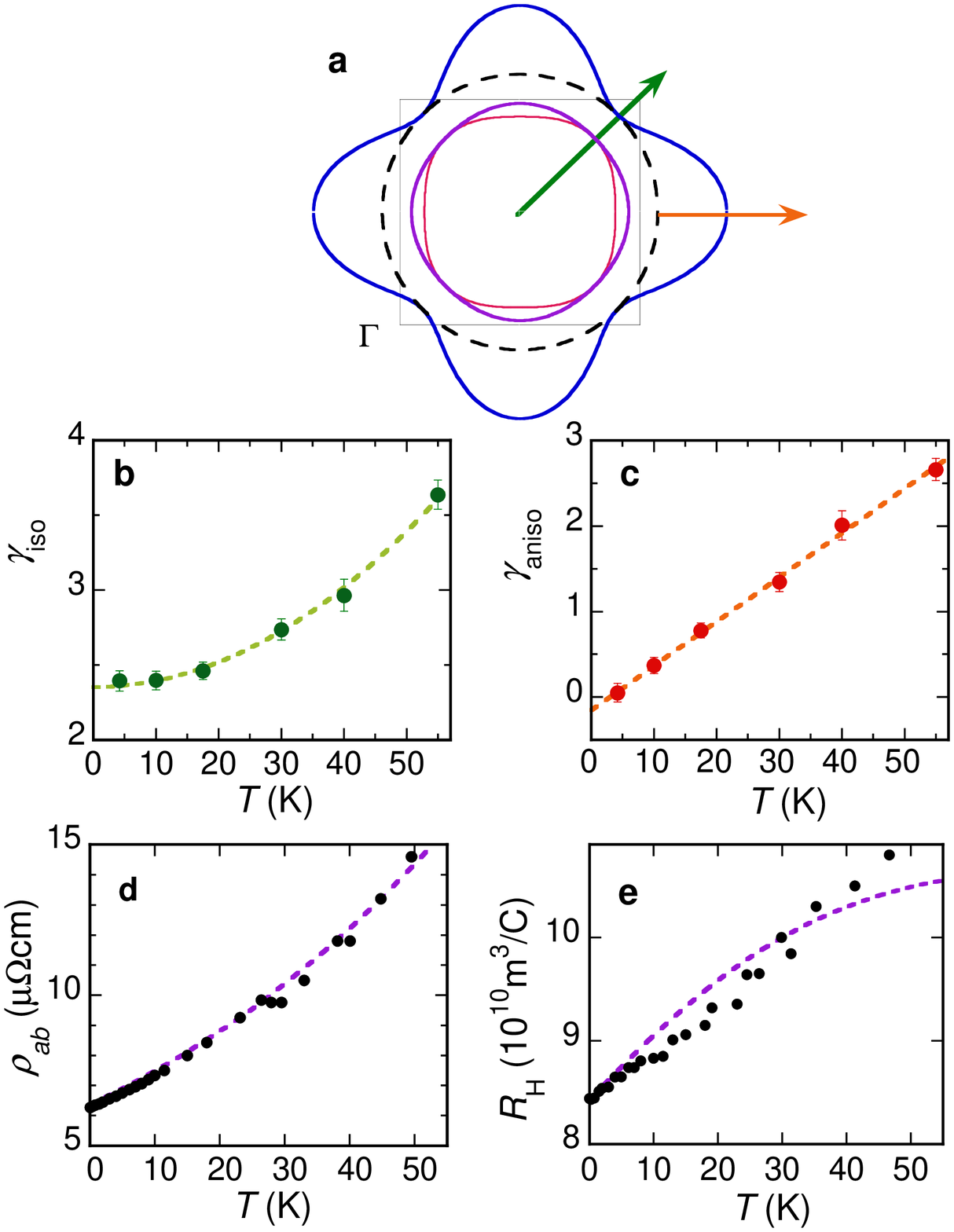} \caption{a) Red curve: schematic 2D
projection of the FS of overdoped Tl2201. Purple curve: schematic representation of the $d$-wave superconducting gap.
Blue curve: geometry of ($\omega_c\tau)^{-1}(\varphi$). Black dashed line: isotropic part of
($\omega_c\tau)^{-1}(\varphi$). b) $T$-dependence of $\gamma_{\rm iso}$, i.e. the isotropic component of
($\omega_c\tau)^{-1}(\varphi$) and sole contribution along the \lq nodal' region indicated by the green arrow in
Fig.~\ref{Figure3}a. The green dashed curve is a fit to $A + BT^2$. c) $T$-dependence of $\gamma_{\rm aniso}$, i.e. the
anisotropic component of ($\omega_c\tau)^{-1}(\varphi$) and the additional contribution that is maximal along the \lq
anti-nodal' direction indicated by the orange arrow in Fig.~\ref{Figure3}a. The orange dashed curve is a fit to $C +
DT$. d) Black circles: $\rho_{ab}(T)$ data for overdoped Tl2201 ($T_c$ = 15 K) extracted from Ref. \cite{Mackenzie96a}.
Purple dashed curve: simulation of $\rho_{ab}(T)$ from parameters extracted from the ADMR analysis. e) Black circles:
$R_{\rm H}(T)$ data for the same crystal \cite{Mackenzie96a}. Purple dashed curve: simulation of $R_{\rm H}(T)$.
Adapted with permission from Nature Physics {\bf 2} 821, figure 2. Copyright 2006 Macmillan Publishers Ltd.}
\label{Figure3}
\end{figure}

In addition to anisotropy in the elastic scattering channel, recent polar ADMR measurements in OD Tl2201 have revealed
anisotropy of a similar symmetry and form in the {\it inelastic} scattering \cite{Majed06}. By incorporating
basal-plane anisotropy in $\omega_c \tau$ into the ADMR analysis, Abdel-Jawad {\it et al.} were able to extract from
the full $T$- and $k$-dependence of $\ell_{ab}$ in heavily OD T2201 ($T_c$ = 15 K) up to 60 K, as shown in Fig.
\ref{Figure3}. The $T$-dependence of the anisotropy was attributed in full to the scattering rate, and from the
resulting fits, the authors were able to conclude that the scattering rate $\Gamma$ contained two components, an
isotropic $T^2$ scattering rate (presumably due to electron-electron scattering) and an anisotropic $T$-linear
component (of unknown origin) that was maximal along ($\pi$, 0). Combining this with the elastic term, the full
expression for the \lq ideal' scattering rate $\Gamma_{\rm ideal}$ in OD cuprates can be written as

\begin{equation}
\Gamma_{\rm ideal}(T, \varphi) = \Gamma_0(\varphi) + \Gamma_1 \cos^2(2\varphi)T + \Gamma_2 T^2 \label{eq2}\\
\end{equation}

where $\Gamma_0(\varphi)$ = $\beta/v_F(\varphi)$, i.e. proportional to the in-plane density of states
\cite{VarmaAbrahams}. Intriguingly, the anisotropic $T$-linear scattering rate has the same anisotropic form as the
$\omega$-linear contribution to Im$\Sigma$ recently uncovered in OP LSCO \cite{Chang07b}. It is also interesting to
note from Fig.~\ref{Figure3}c that in Tl2201, the elastic scattering rate is almost isotropic, in marked contrast to
what is seen in LSCO \cite{Yoshida07, Chang07b}. This difference can be attributed almost entirely to their individual
band structures. In Tl2201, $v_F(\varphi)$ varies by no more than 40$\%$ around the FS \cite{Analytis07}, consistent
with the more rounded FS \cite{Hussey03b, Plate05} and its subsequent displacement away from the saddles. This level of
anisotropy is one order of magnitude smaller than in LSCO. Moreover, the FS in Tl2201 does not possess negative
curvature. Correspondingly, the absolute value of $R_{\rm H}$(0) in OD Tl2201 is found to be in reasonable agreement
with the isotropic-$\ell$ approximation \cite{Mackenzie96a} (see Fig.~\ref{Figure3}e), whilst in OD LSCO, the
combination of negative FS curvature and strong band anisotropy has a significant and highly non-trivial effect on
$R_{\rm H}$(0). Similar detailed considerations may also be relevant to recent high-field Hall measurements on thin
films of the low-$T_c$ cuprate La-doped Bi2201 where a marked drop in $R_{\rm H}$(0) is observed near optimal doping
\cite{Balakirev03}. This discontinuity in the doping dependence of $R_{\rm H}$(0) was originally associated with a
marked change in the FS geometry, perhaps due to a quantum phase transition.

Significantly, the form of $\Gamma_{\rm ideal}(T, \varphi)$ shown in Eqn.~(\ref{eq2}) can also explain, in a
quantitative and self-consistent fashion, the $T$-dependencies of $\rho_{ab}(T)$ and $R_{\rm H}(T)$ in Tl2201 over the
same temperature range as the ADMR measurements and in particular, the persistence of a strong $T$-linear component in
$\rho_{ab}(T)$ to low $T$ reproduced in Fig.~\ref{Figure3}d \cite{Mackenzie96a}. Moreover, as $T$ increases, the growth
in anisotropy in $\Gamma(T, \varphi)$ is sufficient to explain the rise in $R_{\rm H}(T)$, as shown in
Fig.~\ref{Figure3}e. Again, in heavily OD LSCO, the opposite trend seems to occur; the anisotropy starts high then
gradually gets washed out with increasing temperature.

Corroborating evidence for strong basal-plane anisotropy in the {\it transport} scattering rate (with different
$T$-dependencies) comes from Raman measurements \cite{DevereauxReview}. Along the nodal direction ($B_{2g}$ symmetry),
the qp scattering rate is relatively small at all doping levels and shows metallic behaviour at all temperatures.
Around ($\pi$, 0) ($B_{1g}$ symmetry) however, the Raman spectra are highly doping-dependent and in Bi2212, can even
show non-metallic behaviour below $p$ = 0.15 \cite{DevereauxReview}. The ratio of the anti-nodal to nodal scattering
rates $\Gamma(\pi,0)$/$\Gamma(\pi,\pi)$ and their doping dependence, as inferred from ARPES \cite{Zhou04, Plate05,
Yang06}, Raman \cite{DevereauxReview, Tassini04} and transport measurements \cite{Majed07, Hussey03}, are compared in
Fig.~\ref{Figure8}. Whilst all three probes show a similar trend, namely an overall decline in the strength of the
anisotropy with doping, there are some key differences. According to ARPES for example, the nodal/antinodal
quasiparticle anisotropy is seen to vanish \cite{Bogdanov02, Yang06} or even reverse its sign \cite{Zhou04, Plate05}
before superconductivity is suppressed on the OD side, whilst in Raman, the anisotropy collapses abruptly around $p
\sim 0.20$ \cite{DevereauxReview, Tassini04}. By contrast, ADMR analysis on Tl2201 has revealed that the strength of
the anisotropic scattering scales linearly with $T_c$, appearing to extrapolate to zero at the doping level where
superconductivity vanishes \cite{Majed07}. The origin of these discrepancies is not clear at the time of writing. We
note however that in Tl2201 ($T_c \sim$ 30K), the scattering rate deduced from ADMR \cite{Majed07, Hussey96} is more
than one order of magnitude smaller than the corresponding rate derived from ARPES \cite{Plate05}, suggesting that the
transport and quasiparticle lifetimes are in fact distinct.

\begin{figure}
\centering
\includegraphics[width=6.0cm,angle=270,keepaspectratio=true]{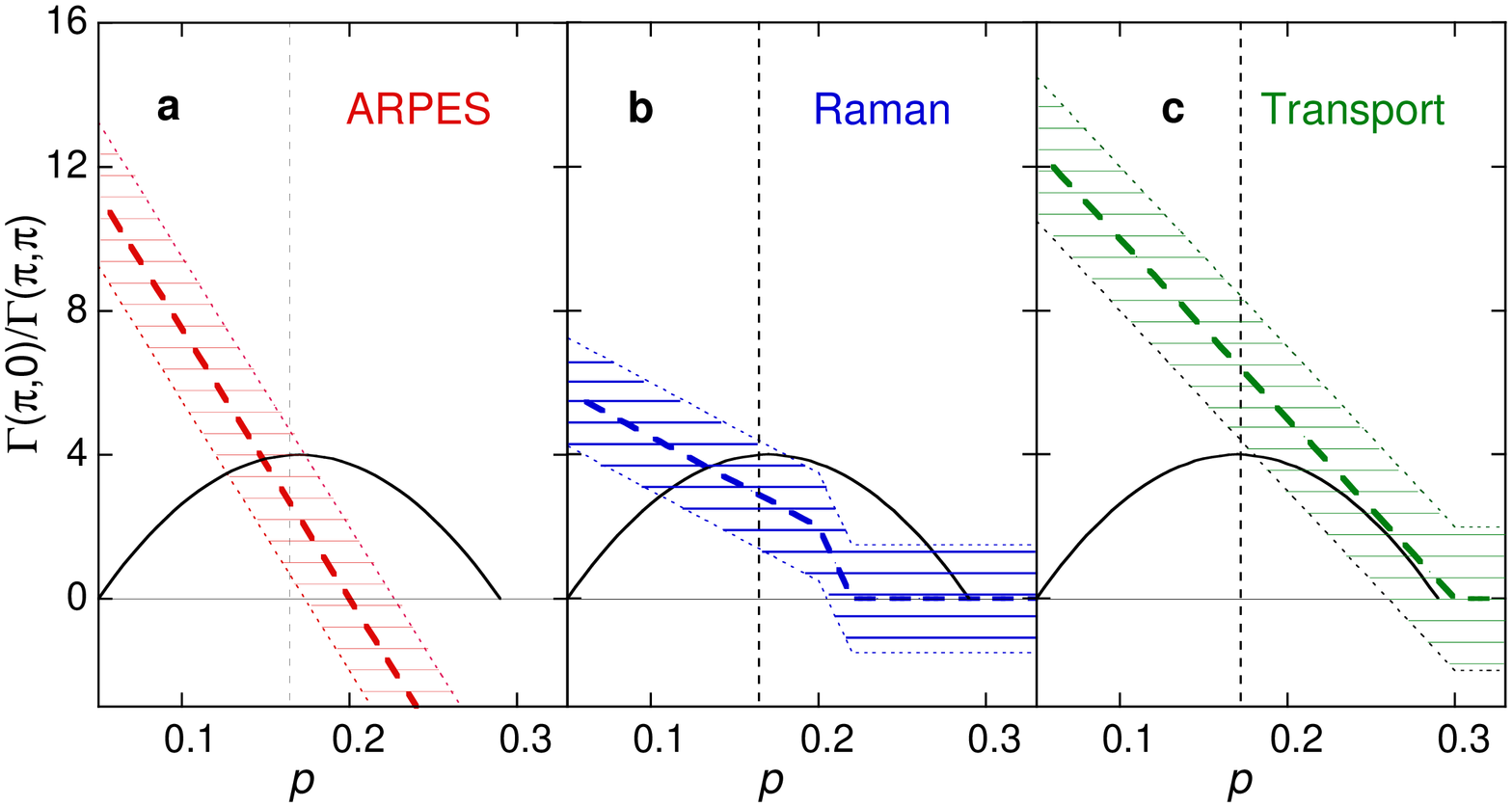}
\caption{Schematic doping evolution of the basal plane anisotropy of the scattering rate, defined as
$\Gamma(\pi,0)$/$\Gamma(\pi,\pi)$, in high-$T_c$ cuprates as revealed by a) ARPES, b) Raman and c) ADMR. The black
solid line represents the superconducting dome whilst the thin dashed line indicates optimal doping.} \label{Figure8}
\end{figure}

But what of the inverse Hall angle? According to the Ong construction, cot$\theta_{\rm H}(T)$ is largely determined by
the behaviour of the longest-lived quasiparticles which in hole-doped cuprates are the nodal quasiparticles near ($\pi,
\pi$). As shown in Fig.~\ref{Figure3}b, the nodal scattering rate is strictly quadratic in temperature \cite{Majed06}.
Provided the anisotropy is sufficiently large, this nodal scattering will dominate the Hall conductivity and account
for the $A +BT^2$ form of cot$\theta_{\rm H}$($T$). Given that Im$\Sigma \propto \omega^2$ along the nodes
\cite{Kordyuk04, Koralek06}, this form of scattering appears to be purely electronic in origin. Whilst the broad energy
range over which this quadratic dependence persists may appear anomalous, similar behaviour is observed in other
correlated oxides such as Sr$_2$RuO$_4$ \cite{Hussey98b}, Sr$_2$RhO$_4$ \cite{Baumberger06} and PrBa$_2$Cu$_4$O$_8$
\cite{McBrien02}. Finally, the reduction in the strength of the anisotropic scattering with doping, highlighted in
Fig.~\ref{Figure8}, can offer an explanation for the softening of the exponent $\alpha$ in the $T$-dependence of
cot$\theta_{\rm H}(T)$ (= $A + BT^\alpha$) in both Bi2201 and Bi2212 (from 2 to 1.65) as one moves across the phase
diagram \cite{AndoMurayama, Konstantinovic00}; as the anisotropy in $\ell$({\bf k}) is reduced, regions of the FS away
from the nodes, where $\ell$({\bf k}) is larger and the $T$-dependence is weaker than $T^2$, begin to contribute to
$\sigma_{xy}$ and thus to cot$\theta_{\rm H}(T)$ \cite{Hussey03}.

\section{Scattering rate saturation}

In recent years, the energy scale over which the qp excitation spectrum can be probed by optical and photoemission
spectroscopies has increased markedly, allowing access to the full spectral response up to the bare (or renormalized)
bandwidth $W$. One striking feature of the high energy response has been the tendency of the single-particle or
particle-particle scattering rate towards saturation. Such behaviour is exemplified by recent state-of-the-art optical
reflectivity data on OP Bi$_2$Sr$_2$Ca$_{0.92}$Y$_{0.08}$Cu$_2$O$_8$ (Y-Bi2212) \cite{vdM03}, analysed using an \lq
extended' or \lq generalized' Drude model that assumes a single Drude component for $\omega < W$ but with a scattering
rate $\Gamma$($T$, $\omega$) and coupling constant $\lambda$($T$, $\omega$) that are frequency-dependent.

As shown in Fig.~\ref{Figure9}a, $\Gamma$($T, \omega$) initially grows linearly with frequency but above 3000
cm$^{-1}$, it starts to deviate from linearity, tending to a constant value $\Gamma_{\rm max} \sim$ 4000 cm$^{-1}$
($\sim$ 0.5 eV). At first sight, such \lq saturation' in $\Gamma$($T$, $\omega$) appears suggestive of strong coupling
to bosons, though the high frequency at which saturation sets in ($\sim$ 4000 cm$^{-1}$) implies that these are not
phonons \cite{Shulga91}. Norman and Chubukov \cite{NormanChubukov} recently argued that a model based on coupling to a
broad spectrum of spin fluctuations, extending out to 0.3 eV, captures most of the essential features of the data in
Ref. \cite{vdM03}, although a gapped MFL model was also found to work reasonably well. Similarly, Hwang {\it et al.}
extracted Eliashberg-type functions from optical conductivity data on UD YBCO \cite{Hwang06} and OP LSCO
\cite{Hwang07b} and showed them to be consistent with the spectrum of spin fluctuations obtained independently from INS
measurements \cite{Stock05, Vignolle07}. In each case however, neglect of the basal-plane anisotropy in $\Gamma$($T$,
$\omega$) in the modelling may have compromised the analysis significantly \cite{Hussey06}.

\begin{figure}
\centering
\includegraphics[width=5.5cm,angle=270,keepaspectratio=true]{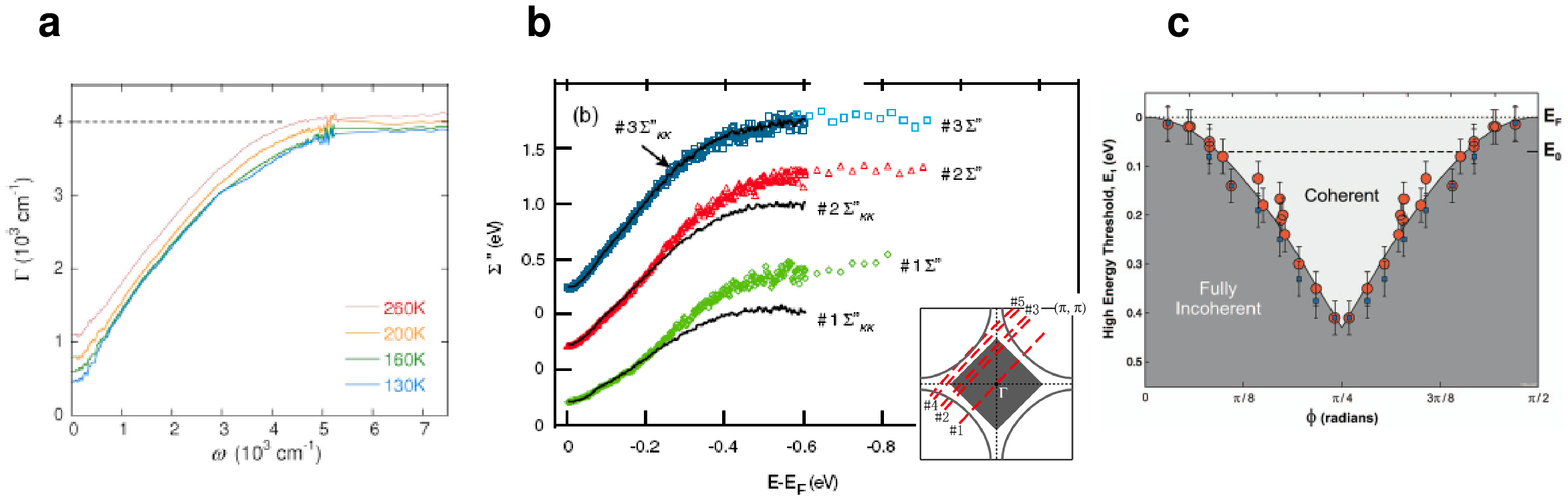}
\caption{Scattering rate of a) OP Bi2212 \cite{vdM03} and b) OD Bi2201 \cite{Xie07} determined by optical conductivity
and ARPES measurements respectively. In a), the different coloured curves represent $\Gamma(\omega)$ determined for
several temperatures using an extended Drude analysis of reflectivity data. The dashed line is the saturation value of
$\Gamma(\omega)$. Reprinted with kind permission from \cite{Hussey06}, figure 1. Copyright 2006 by the American
Physical Society In b), the diamonds, triangles and squares represent Im$\Sigma(\omega)$ for different cuts across the
in-plane Fermi surface (shown as dashed red lines in the inset) determined from the widths of the momentum distribution
curves. The solid lines labelled $\Sigma^{''}_{\rm KK}$ are equivalent curves derived from Kramers-Kronig
transformation of Re$\Sigma(\omega)$. Reprinted with kind permission from \cite{Xie07}, figure 4b. Copyright 2007 by
the American Physical Society. c) Onset of saturation as a function of azimuthal angle in OP LSCO. 0 and $\pi/2$ refer
to those regions closest to the saddle points \cite{Chang07a}. Reprinted with kind permission from \cite{Chang07a},
figure 3. Copyright 2007 by the American Physical Society.} \label{Figure9}
\end{figure}

One aspect of the data in Fig. \ref{Figure9}a that is at odds with the standard picture of electron-boson coupling is
its $T$-dependence. According to the original Allen formalism \cite{Shulga91, Allen71}, saturation of
$\Gamma$($\omega$) sets in at progressively higher frequencies as $T$ is raised, more so still if the bosonic response
were to broaden and shift to higher frequencies, as is expected if the strongest coupling is to antiferromagnetic spin
fluctuations \cite{Hwang07b}. The data do {\it not} show this tendency; if anything, $\Gamma$($\omega$) saturates at a
{\it lower} frequency as $T$ increases. This counter trend in $\Gamma$($T$, $\omega$) is seen particularly clearly in
the optical response of UD Ca$_{2-x}$Na$_x$CuO$_2$Cl$_2$ \cite{Waku}. Hwang {\it et al.} proposed a way around this
problem by combining their Eliashberg-type analysis with an frequency-dependent density of states that served to mimic
the pseudogap in UD YBCO \cite{Hwang06} though yet again, no account was made of its angular dependence.

Saturation is also seen in the single-particle scattering rate Im$\Sigma$, an example of which is shown in Fig.
\ref{Figure9}b for OD Bi2201 \cite{Xie07}. As with the optical scattering rate, saturation does not set in until
extremely high energies, of order 0.5 eV (4000 cm$^{-1}$), and attains values of order 1 eV ($\sim W$) that suggests
the development of incoherent excitations \cite{Xie07}. Similar observations have now been reported for OP Bi2212
\cite{Valla07} and OP LSCO \cite{Chang07a} . Notably in OP LSCO, Im$\Sigma(\omega)$ appears to saturate at different
points on the FS at different energies, in line with the strong basal-plane anisotropy in Im$\Sigma(\omega,\varphi)$
\cite{Chang07a, Chang07b}. As shown in Fig. \ref{Figure9}c, Im$\Sigma(\omega)$ near ($\pi$, 0) is found to saturate
close to the Fermi level, whilst along the nodes, coherence survives up to 0.4 eV. This is reminiscent of the evolution
of Im$\Sigma$ with temperature in Bi2212 where anti-nodal states become incoherent above $T_c$ \cite{Kaminski03},
whilst those along the diagonals remain coherent up to much higher temperatures \cite{Kordyuk04, Koralek06}.

An alternative way to interpret this coherent/incoherent crossover and the onset of saturation in $\Gamma$($\omega$) is
to invoke the Mott-Ioffe-Regel (MIR) limit for coherent charge propagation. The MIR criterion states that the electron
mean-free-path $\ell$ has a lower limit of order the interatomic spacing $a$ (or to put it another way, $\Gamma$ can
never exceed the bare bandwidth $W$). Beyond that point, the concept of carrier velocity is lost and all coherent
quasiparticle motion vanishes. Such a threshold is seen, for example, in metals exhibiting resistivity saturation,
where the saturation value is found to be consistent with $\ell$ = $a$ \cite{Gunnarsson, Hussey04}. For Y-Bi2212, one
can estimate $\Gamma_{\rm max}$ = $\langle v_F \rangle$/$a \sim$ 4500 cm$^{-1}$ using parameters independently
extracted from both optics \cite{vdM03} and ARPES \cite{Kaminski05}. Comparison with Fig. \ref{Figure9}a suggests that
the saturation value of $\Gamma$($T$, $\omega$) is indeed comparable with the MIR limit as defined.

Puzzlingly, saturation in $\Gamma$($\omega$) at or near the MIR limit does not appear to be manifest in the dc
transport properties where, in OP LSCO for example, $\rho_{ab}$($T$) grows approximately linearly with $T$ up to 1000 K
reaching values of the order 1 - 10 m$\Omega$cm that correspond to $\ell \ll a$ \cite{GurvitchFiory}. Detailed analysis
of optical conductivity data in cuprates and in other so-called \lq bad metals' however has shown that as the MIR limit
is attained, loss of coherence is identified by a suppression of low-frequency spectral weight and the development of a
non-Drude optical response \cite{Hussey04}. The lost weight is transferred to energies $\omega > W$ ($\sim$ 1 eV) and
the optical sum-rule is only fulfilled at much higher energies of 2 - 3 eV \cite{Merino00, Takenaka03}. This collapse
of the zero-frequency collective mode implies that continuation of the positive slope of $\rho_{ab}(T)$ well into the
incoherent regime has nothing to do with an unbounded escalation of the scattering rate. It also brings into question
association of saturation in $\Gamma(\omega)$ with bosonic mode coupling, since there one might expect the scattering
rate to continue to grow and the Drude response continue to broaden. This is not what is observed.

\begin{figure}
\centering
\includegraphics[width=8.5cm,angle=270,keepaspectratio=true]{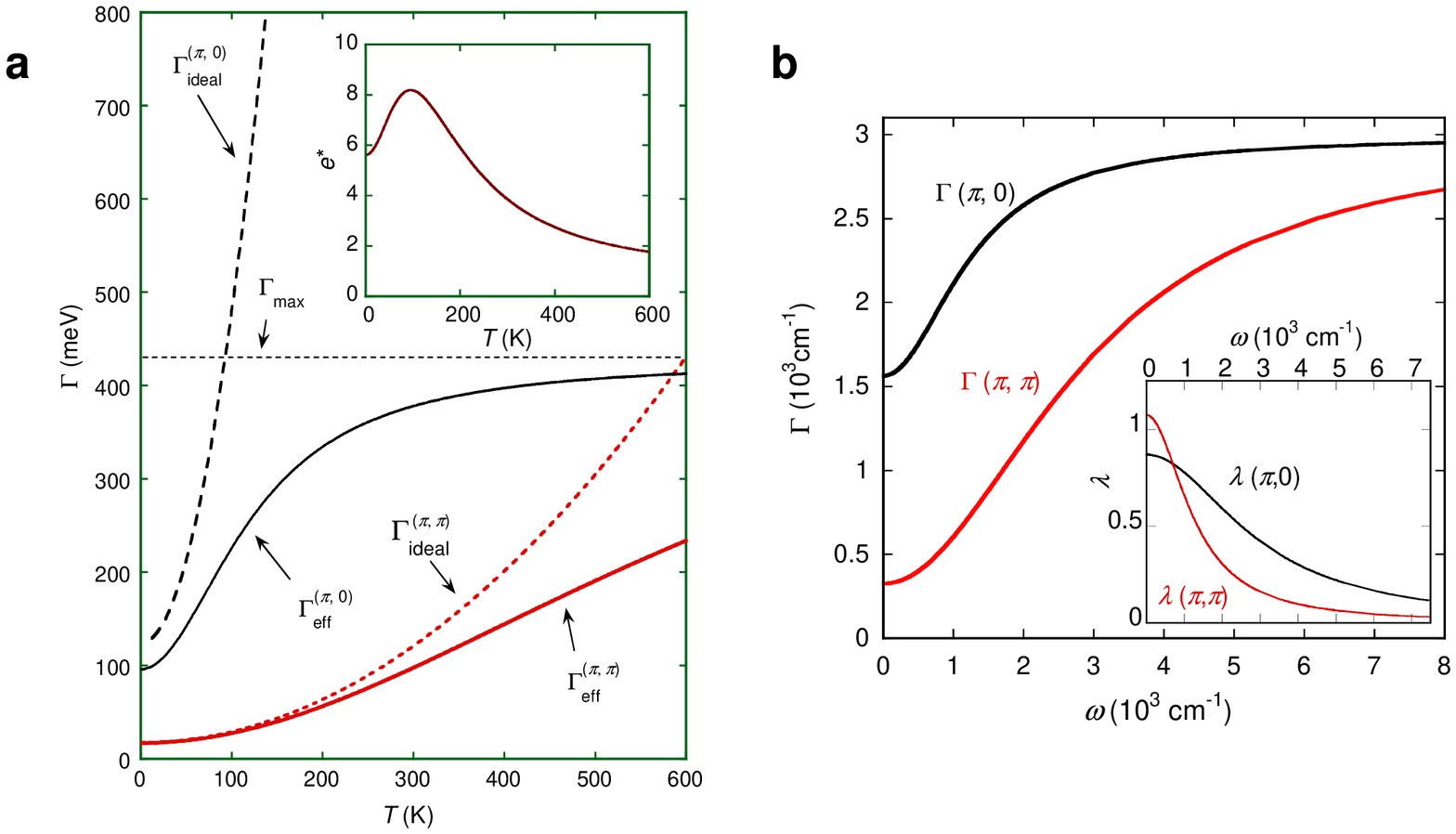}
\caption{Left panel: $T$-dependence of the ideal and effective scattering rates for ($\pi$, 0) and ($\pi$, $\pi$) for
OP Tl2201. Inset: $T$-dependence of the effective anisotropy $e$* = $\Gamma_{\rm{eff}}^{(\pi,
0)}$/$\Gamma_{\rm{eff}}^{(\pi, \pi)}$ - 1 for the same parameterization. Adapted from \cite{Hussey03}. Copyright 2003
by Springer. Right panel: $\Gamma_{\rm eff}$($\omega$) along ($\pi$, $\pi$) (red) and ($\pi$, 0) (black) at $T$ = 200K
for OP Y-Bi2212. Inset: Corresponding $\lambda_{\rm eff}$($\omega$) for the same two orientations, obtained via the
appropriate Kramers-Kronig transformation. Reprinted with kind permission from \cite{Hussey06}, figure 2. Copyright
2006 by the American Physical Society.} \label{ASS}
\end{figure}

So how does saturation in $\Gamma(\omega)$ at high energies impact on the dc transport properties? In metals that
exhibit resistivity saturation, the form of the resistivity is extremely well captured by the so-called \lq
parallel-resistor' model \cite{Wiesmann77}, 1/$\rho(T)$ = 1/$\rho_{\rm ideal}(T)$ + 1/$\rho_{\rm max}$, where
$\rho_{\rm ideal}$ is the ideal resistivity (i.e. in the absence of saturation, typically the Bloch-Gr\"{u}neisen
resistivity) that is shunted by a large saturation resistivity $\rho_{\rm max}$ corresponding to $\ell = a$. Whilst the
origin of this behaviour is still not understood, one can readily assume that the parallel-resistor form of the
resistivity in such metals reflects that of the scattering rate which one can easily modify to include anisotropy
\cite{Hussey03} by defining an \lq effective' scattering rate $\Gamma_{\rm eff}$ as

\begin{equation}
1/\Gamma_{\rm eff}(T, \varphi) = 1/\Gamma_{\rm ideal}(T, \varphi) + 1/\Gamma_{\rm max} \label{eq3}\\
\end{equation}

Note that this formula implies scattering rates adding in parallel rather than different conduction channels. Use of
Eqn.~(\ref{eq3}) acknowledges that the MIR limit is manifest at all temperatures and, by extension, all energies below
the bandwidth, hence its potential impact on the dc transport. In optimally and overdoped HTC, $ 1/\Gamma_{\rm
ideal}(T, \varphi)$ is given by Eqn.~(\ref{eq2}). A schematic demonstration of the role of Eqn.~(\ref{eq3}) is shown in
Fig.~\ref{ASS}a where the $T$-dependencies of $\Gamma_{\rm eff}$($T, \varphi$) for the two key momentum directions
($\pi$, 0) and ($\pi$, $\pi$) are plotted. For illustrative purposes, $\Gamma_1$ = 0 and the anisotropy parameters used
in this simulation are those applied originally to OP Tl2201 to produce a $T$-linear resistivity, an inverse Hall angle
cot$\theta_{\rm H} \sim A + BT^2$ and a modified Kohler's rule of the correct magnitude \cite{Hussey03}. Along ($\pi$,
0), $\Gamma_{\rm{ideal}}$($T$) reaches $\Gamma_{\rm max}$ at $T_c$ $\sim$ 90 K, at which point, the anti-nodal
quasi-particles start to lose coherence. Due to the presence of the \lq shunt' $\Gamma_{\rm max}$, however,
$\Gamma_{\rm eff}$($T$) is always smaller than $\Gamma_{\rm ideal}$($T$) and approaches saturation much more gradually.
For the quasi-particles at ($\pi$, $\pi$) on the other hand, $\Gamma_{\rm ideal}$($T$) reaches $\Gamma_{\rm max}$ at a
much higher temperature ($T \sim$ 600K). Hence, at the nodal points, well-defined coherent quasi-particles exist at all
relevant temperatures. Note that $\Gamma_{\rm eff }$($T$) at ($\pi$, $\pi$) is quasi-linear over a wide temperature
range. Within this model, the onset of $T$-linear resistivity above $T_{\rm coh}$ (see Fig.~\ref{Figure1}) coincides
with the temperature at which $\Gamma_{\rm ideal}^{(\pi,0)}$ exceeds $\Gamma_{\rm max}$, in excellent agreement with
what has been inferred from ARPES measurements on OP Bi2212 \cite{Kaminski03, Kordyuk04, Koralek06}.

Correspondingly, the \lq effective' anisotropy factor $e$* (= $\Gamma_{\rm eff}^{(\pi, 0)}$/$\Gamma_{\rm eff}^{(\pi,
\pi)}$ - 1) also reaches a maximum value at this temperature (see inset to Fig.~\ref{ASS}a). The overall effect is a
transport scattering rate $\Gamma_{\rm eff}$($T, \varphi$) whose anisotropy initially grows with increasing $T$ but
then gradually becomes more isotropic at higher temperatures as the scattering rate at different regions of the FS
tends towards saturation. In this way, the $T$-dependence of $e$*($T$) (Fig.~\ref{ASS}a) mimics that of the Hall
coefficient $R_{\rm H}$($T$) in HTC; another example of saturation playing a key role in a dc transport property.
Finally, with increasing doping (away from optimal doping), the relative anisotropy decreases gradually to zero
(Fig.~\ref{Figure8}) as one approaches the non-superconducting normal metallic region \cite{Majed07, Hussey03,
Yosuf02}, leading to a lessening of the $T$-dependence of $R_{\rm H}(T)$ as observed experimentally \cite{Kubo91}.

It is relatively straightforward to extend this idea into the frequency domain \cite{Hussey06}. Fig.~\ref{ASS}b shows
the evaluated frequency-dependent scattering rate $\Gamma_{\rm eff}$($\omega$) along ($\pi$, $\pi$) (red line) and
($\pi$, 0) (black line) used to fit the optical conductivity data on Y-Bi2212 at $T$ = 200K \cite{vdM03}. The inset
shows the corresponding mass enhancement factors $\lambda_{\rm eff}$($\omega$) obtained via the appropriate
Kramers-Kronig transformation. (Again for simplicity, $\Gamma_1$ = 0 here). The anisotropy in $\lambda_{\rm eff}$ is
considerably weaker than in $\Gamma_{\rm eff}$, highlighting the surprisingly large effect of saturation on the mass
enhancement, even at the dc limit. Indeed, suppression of $\lambda(0)$, the zero-frequency coupling constant, becomes
stronger as $\Gamma_{\rm ideal}(\omega = 0)$ approaches $\Gamma_{\rm max}$ \cite{Hussey05}; a feature that may help to
explain why mass enhancement in HTC does not diverge near the Mott transition.

As described above, this dichotomy in the nodal and anti-nodal qp states has been intimated by a variety of different
experiments, including ARPES \cite{Chang07a, Kaminski03, Koralek06}, Raman \cite{Tassini04}, Hall effect
\cite{Fruchter07} and even STM \cite{McElroy05}. Whilst it appears difficult to differentiate between total incoherence
(suppression of spectral weight) and saturation of the scattering rate at the MIR limit, the inclusion of a maximal
scattering rate in the analysis appears particularly instructive and phenomenologically at least, it can help to
explain a wide range of physical observables. Its overall effect is to reduce the anisotropy between the nodal and
anti-nodal states, thereby masking to some extent the intrinsic anisotropy created by whatever physical process is
responsible. Moreover, by invoking the \lq parallel-resistor' model, the saturation of the $T$- or $\omega$-dependent
scattering rate at the MIR limit is seen to influence the transport behavior in cuprates over a very wide energy scale,
even down to the dc limit, and to acknowledge its presence may yet turn out to be a key step in the development of a
coherent description of the charge dynamics.

\section{Discussion and conclusions}

In this review article, I have highlighted a number of recent experimental results which shed new light on the cuprate
transport problem and have identified those ingredients I believe are necessary for a complete phenomenological
description of the anomalous normal state transport properties of optimally doped and overdoped high-$T_c$ cuprates.
These ingredients are summarized in Eqn. (1-3) in the main body of the article; namely the tight-binding band
structure, basal-plane anisotropy in both the elastic and the inelastic $T$-linear scattering channels, an isotropic
electron-electron scattering term and finally, saturation of all scattering at or near the Mott-Ioffe-Regel limit.

In order to investigate to what extent this phenomenology captures the overall physical behaviour in HTC, I show here a
simple one-parameter scaling of the doping evolution of $\rho_{ab}$($T,p$) and $R_{\rm H}$($T,p$) in Tl2201. In this
simulation, the FS parameters are fixed by the ADMR measurements on OD Tl2201 ($T_c$ = 15 K, see Fig. \ref{Figure1}c
\cite{Majed07}) and scaled for each $p$-value by the empirical relation $T_c$/$T_c^{\rm max} = 1 - 82.6(p-0.16)^2$ with
$T_c^{\rm max}$ = 92K \cite{Presland91}. Secondly, $\Gamma_{\rm ideal}(\varphi, T)$ is expressed using Eqn. (2) with
$\Gamma_0$ and $\Gamma_2$ assumed to be doping (and $k$-)independent and fixed by the values determined for the $T_c$ =
15K sample \cite{Majed07}. Hence the only parameter that is allowed to vary is the strength of the anisotropic
$T$-linear term $\Gamma_1$ = $\Gamma_1$(15K) $\times$ $T_c(p)/15$. The anticipated return to isotropic scattering at
high $T$ is simulated by inclusion (in parallel) of a maximum scattering rate $\Gamma_{\rm max}$ (= $\langle v_F
\rangle/a$) in accord with the MIR limit (Eqn. (3)). The corresponding $\Gamma_{\rm eff}(T, \varphi)$ is then inserted,
along with the FS parameters, into the appropriate expressions for $\rho_{ab}$ and $R_{\rm H}$ derived using the
Jones-Zener form of the Boltzmann equation for a quasi-2D FS \cite{Hussey03}.

\begin{figure}
\centering
\includegraphics[width=5.5cm,angle=270,keepaspectratio=true]{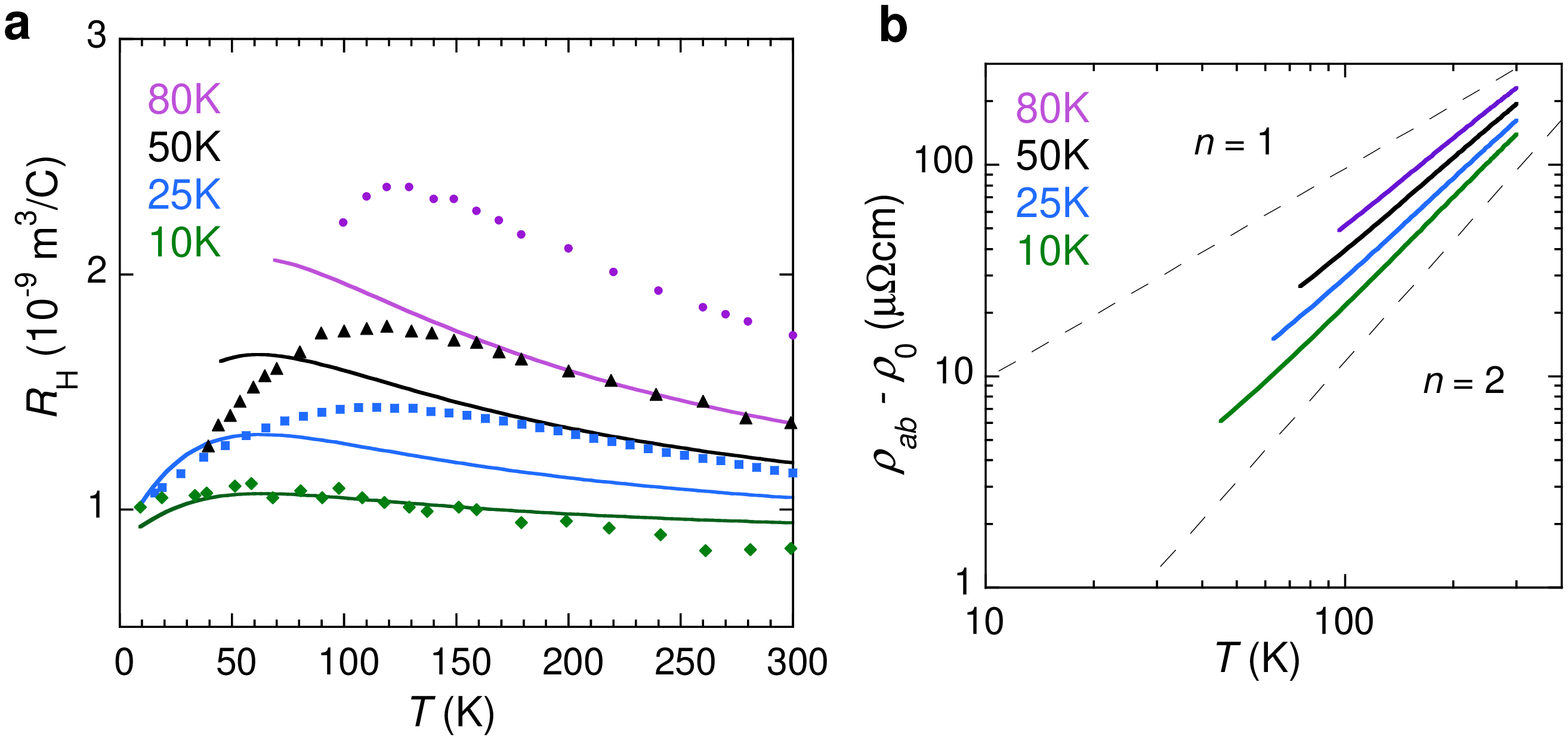}
\caption{a) $R_{\rm H}(T)$ simulations for Tl2201 with $T_c$ values of 10, 25, 50 and 80K (solid lines) together with
published data for $10$K (green diamonds \cite{Manako92}), 25K (blue squares \cite{Hussey96})and 50K (black triangles
\cite{Manako92}) single crystals plus polycrystalline data for $T_c$ = 81K (purple circles, \cite{Kubo91}). b)
Simulation of $\rho_{ab}(T) - \rho_{ab}(T=0)$ (on a log-log scale) using the same parameterization. Adapted with kind
permission from \cite{Majed07}, figure 3. Copyright 2007 by the American Physical Society.} \label{Fits}
\end{figure}

The resultant simulations for $R_{\rm H}$($T,p$) and $\rho_{ab}$($T,p$) are presented in Fig.~\ref{Fits} for $T_c$ =
10, 25, 50 and 80 K. Despite there being no free parameters in the modelling ($\Gamma_1$ is essentially fixed by
$T_c$), the $R_{\rm H}$($T$) plots show good qualitative agreement with the published data (solid symbols,
\cite{Manako92, Hussey96, Kubo91}). The $p$-dependence of $\rho_{ab}(T)$ (= $\rho_0 + \nu T^n$) is also found to be
reasonably consistent with experiment, with the exponent $n$ evolving smoothly from 2 at the SC/non-SC boundary towards
unity as one approaches optimal doping \cite{Kubo91}. The fact that our simulations capture the marked ($\times 2$)
increase in the magnitude of $R_{\rm H}$ (and $\rho_{ab}$) for what are relatively small ($\sim 5 \%$) changes in
carrier concentration, suggests that anisotropic scattering is indeed a dominant contributor to the $T$- and
$p$-dependence of $R_{\rm H}$ and $\rho_{ab}$ across the OD regime, whilst the {\it monotonic} variation of $R_{\rm
H}$($T$, $p$) argues against any sign reversal of the scattering rate anisotropy near $p$ = 0.2 as inferred from ARPES
\cite{Zhou04, Plate05}. Discrepancies between the fitting and the model seem to grow stronger as one progressively
lowers the doping. These might be attributable to the gradual emergence of current vertex corrections (also known as
\lq backflow') that act to modify the direction of the effective quasiparticle velocity (and thus the direction of the
total current), leading to exaggeration of the curvature in $\ell$($k$) and hence a stronger renormalization of $R_{\rm
H}$ \cite{Kontani06}. Since interlayer transport in cuprates is believed to be determined by a product of {\it
single-particle} spectral functions on adjacent planes \cite{Sandemann01}, it is possible that parameters obtained by
ADMR, an {\it interlayer} tunneling measurement, do not include these corrections.

Such details notwithstanding, the overall experimental situation does appear to support models in which anisotropic
$T$-linear scattering, in conjunction with the Fermi surface curvature and band anisotropy \cite{Ong91}, is primarily
responsible for the $T$-dependence of $\rho_{ab}(T)$ and $R_{\rm H}(T)$. The origin of the anisotropic scattering is
not known at present. Interactions with a bosonic mode are one obvious candidate, though given the strong angle and
doping dependence, presumably not phonons. Whilst phonons can give rise to anisotropic scattering, the behaviour of
$\Gamma(\omega)$, extracted from extended Drude analysis of the in-plane optical conductivity, as described above, is
inconsistent with an electron-boson scattering response due to phonons \cite{vdM03}. Moreover, it has proved
problematic to explain the quadratic $T$-dependence of the inverse Hall angle cot$\theta_{\rm H}(T)$ in a scenario
based solely on electron-phonon scattering.  More likely candidates include spin \cite{Carrington92, MonthouxPines},
charge (stripe) \cite{Castellani95} or $d$-wave pairing fluctuations \cite{IoffeMillis}, all of which disappear with
superconductivity on the OD side \cite{IoffeMillis, Wakimoto04, Reznik06}. The preservation of the $T$-linear
scattering rate to low $T$ however would seem to imply a vanishingly small energy scale for such fluctuations,
characteristic of proximity to a quantum critical point. This feature of the scattering is perhaps more consistent with
MFL phenomenology, particularly since a similar $\omega$-linear dependence has also been seen in Im$\Sigma$ (in LSCO)
\cite{Chang07b}. Its vanishing at the nodes and the coexistence of a ubiquitous $T^2$ dependence however represent
significant challenges to this scenario that have yet to be addressed. In the original two-lifetime picture meanwhile,
no basal-plane anisotropy has ever been considered. More recently, the transport scattering rate close to a 2D
Pomeranchuk instability was shown to have a form identical to that observed in OD Tl2201 \cite{Dell'Anna07}. In such a
scenario however, the magnitude of the anisotropy should grow, rather than diminish, with doping as the vHs is
approached on the OD side. Finally, real-space (correlated) electronic inhomogeneity \cite{McElroy05} cannot be
excluded as a possible origin of the {\bf k}-space anisotropy, though as yet, no measurements have been performed on
heavily OD non-SC cuprates to establish any possible link between inhomogeneity and superconductivity. Thus, at the
time of writing, it appears that none of the existing models are wholly consistent with the form derived for
$\ell$({\bf k}) and clearly, more theoretical input is required.

Extension of the analysis of ARPES and optical conductivity measurements to high energies has revealed the tendency
towards saturation in both the single-particle and the particle-particle scattering rates with an onset that occurs at
different temperatures and/or energies depending on the location in {\bf k}-space. Saturation generally masks the
intrinsic nature of $\Gamma$($\omega$) in many of the physical properties that are measured and this may explain why it
has taken the community so long to reach a consensus on the various interactions and scattering mechanisms that
influence the self-energy of the in-plane anti-nodal quasiparticles in cuprates and which ultimately, may drive high
temperature superconductivity. Measurements that focused on the heavily overdoped region of the phase diagram, where
saturation is less of an issue, have been able to reveal for the first time the precise form of the additional
scattering rate that is maximal at the anti-nodes and increases in intensity as the doping level is reduced
\cite{Majed06, Majed07}. What intrinsic form this anisotropic scattering takes in the underdoped region may never be
known, particularly since the pseudogap itself may act to suppress scattering in a highly non-trivial and anisotropic
fashion and require a detailed, self-consistent, microscopic formalism to address. On the experimental side, more
systematic $k$-dependent transport and spectroscopic studies in this region of the phase diagram would be highly
beneficial. It would also be helpful to search for universality amongst the different classes of HTC compounds.
Although similar phenomenology in the transport properties has been reported in Tl2201, Bi2212 and Bi2201, and
differences in $R_{\rm H}(T)$ in LSCO can be explained largely by differences in its Fermiology, there are still
outstanding issues, for example, the very strong $T$-dependence of $R_{\rm H}(T)$ in YBCO, that need to be addressed.

But I do not wish to end on a pessimistic note. Indeed, as chronicled throughout this review, the last few years have
witnessed remarkable progress as experimentalists have developed more guile and delved deeper into the mysteries of
HTC. Almost in celebration of its Coming of Age, very recent measurements involving pulsed high magnetic fields have
uncovered a wealth of remarkable and previously unforeseen features of the low $T$ transport behaviour of UD YBCO, such
as quantum oscillations \cite{Doiron-Leyraud07, Yelland07, Bangura07}, negative Hall coefficient \cite{LeBoeuf07} and
metallic $T^2$ resistivity \cite{Rullier07, Proust07}, all of which have sparked renewed debate within the community.
These important developments have provided new insight into the nature of the Fermiology and qp states in underdoped
cuprates and will no doubt feature strongly in future interpretations of the normal state transport properties of HTC.
I look forward to such developments with interest.

\section{Acknowledgements}

The author would like to acknowledge M. Abdel-Jawad, J. G. Analytis, L. Balicas, A. F. Bangura, A. Carrington, R. A.
Cooper, M. M. J. French and A. Narduzzo for contributing to this work and helping the author to formulate many of the
ideas presented. This work was supported by EPSRC (UK).

\end{document}